\documentclass[manuscript]{aastex}

\newcommand{\teff}{\mbox{$T_{\rm eff}$}}
\newcommand{\logg}{\mbox{$\log g$}}
\newcommand{\vsini}{\mbox{$v \sin i$}}
\newcommand{\mictrb}{\mbox{$\xi_{\rm t}$}}
\newcommand{\ftot}{\mbox{$f_{\oplus}$}}
\newcommand{\kms}{\mbox{km\,s$^{-1}$}}
\def\sun{\hbox{$_\odot$}}

\shorttitle{Exotic Nucleus of SuWt~2}
\shortauthors{Exter et al.}

\begin{document}

\title{The Exotic Eclipsing Nucleus of the Ring Planetary Nebula
SuWt~2\altaffilmark{1}}

\author{
K.\ Exter,\altaffilmark{2}
Howard E. Bond,\altaffilmark{3,4}
K.\ G.\ Stassun\altaffilmark{5,6}
B.\ Smalley,\altaffilmark{7}
P.\ F.\ L.\ Maxted,\altaffilmark{7}
and
D.\ L.\ Pollacco\altaffilmark{8}
}

\altaffiltext{1}  
{Based in part on observations obtained   with the SMARTS Consortium 1.3- and
1.5-m telescopes located at Cerro Tololo Inter-American Observatory, Chile, and
with ESO Telescopes at the La Silla Observatory}

\altaffiltext{2}
{Space Telescope Science Institute, 3700 San Martin Dr., Baltimore, MD 21218,
USA\null. Current address: Institute voor Sterrenkunde, Katholieke Universiteit
Leuven, Leuven, Belgium; katrina@ster.kuleuven.be}

\altaffiltext{3}
{Space Telescope Science Institute, 3700 San Martin Dr., Baltimore, MD
21218, USA; bond@stsci.edu}

\altaffiltext{4}
{Visiting astronomer, Cerro Tololo Inter-American Observatory, National Optical
Astronomy Observatory, which is operated by the Association of Universities for
Research in Astronomy, under contract with the National Science Foundation.
Guest Observer with the {\it International Ultraviolet Explorer}, operated by
the Goddard Space Flight Center, National Aeronautics and Space Administration.}

\altaffiltext{5}
{Dept.\ of Physics and Astronomy, Vanderbilt University, Nashville, TN 32735,
USA}

\altaffiltext{6}
{Department of Physics, Fisk University, 1000 17th Ave.\ N, Nashville, TN 37208,
USA}

\altaffiltext{7}
{Astrophysics Group, Chemistry \& Physics, Keele University, Staffordshire, ST5
5BG, UK}

\altaffiltext{8}
{Queen's University Belfast, Belfast, UK}

\begin{abstract}

SuWt~2 is a planetary nebula (PN) consisting of a bright ionized thin ring seen
nearly edge-on, with much fainter bipolar lobes extending perpendicularly to the
ring. It has a bright (12th-mag) central star, too cool to ionize the PN, which
we discovered in the early 1990's to be an eclipsing binary. Although it was
anticipated that there would also be an optically faint, hot, ionizing star in
the system, a spectrum from the {\it International Ultraviolet Explorer\/} ({\it
IUE}) did not reveal a UV source. We present extensive ground-based photometry
and spectroscopy of the central binary collected over the ensuing two decades,
resulting in the determination that the orbital period of the eclipsing pair is
4.9~days, and that it consists of two nearly identical A1\,V stars, each of mass
$\sim$$2.7\,M_\odot$\null. 

The physical parameters of the A~stars, combined with evolutionary tracks, show
that both are in the short-lived ``blue-hook'' evolutionary phase that occurs
between the main sequence and the Hertzsprung gap, and that the age of the
system is about 520~Myr.  One puzzle is that the stars' rotational velocities
are different from each other, and considerably slower than synchronous with the
orbital period. It is possible that the center-of-mass velocity of the eclipsing
pair is  varying with time, suggesting that there is an unseen third orbiting
body in the system. We propose a scenario in which the system began as a
hierarchical triple, consisting of a $\sim$$2.9\,M_\odot$ star orbiting the
close pair of A~stars. Upon reaching the AGB stage, the primary engulfed the
pair into a common envelope, leading to a rapid contraction of the orbit and
catastrophic ejection of the envelope into the orbital plane. In this picture,
the exposed core of the initial primary is now a white dwarf of
$\sim$$0.7\,M_\odot$, orbiting the eclipsing pair, which has already cooled
below the detectability possible by {\it IUE\/} at our derived distance of
2.3~kpc and a reddening of $E(B-V)=0.40$. The SuWt~2 system may be destined to
perish as a Type~Ia supernova.

\end{abstract}


\section{Introduction}

Near the ends of their lives, many, or possibly all, low- and intermediate-mass
stars eject the tenuous outer layers they had developed on the asymptotic giant
branch (AGB), evolve rapidly to high surface temperature, and ionize the
surrounding ejecta, producing a planetary nebula (PN)\null. The detailed
processes of nebular ejection remain uncertain; however, over the past few
decades, it has become clear that binary-star interactions play a role in the
production of a significant fraction of PNe.  The evidence supporting this view
includes (1)~approximately 10--20\% of planetary-nebula nuclei (PNNi) are
photometrically variable, and sometimes eclipsing, binaries with short orbital
periods of hours to a few days (Bond \& Livio 1990; Bond 2000; Miszalski et al.\
2008a, 2009; and references therein); (2)~among a small sample of bright PNNi,
nearly all showed short-timescale radial-velocity variability (De~Marco et al.\
2004), suggestive of membership in close binaries; and (3)~the prevalence of
non-spherical morphologies among PNe. While an axisymmetric morphology can
easily result if ejection is shaped, or indeed directly caused by, binary
interactions, alternative models that have successfully recreated these
morphologies include magnetized winds and rapid stellar rotation (cf.\
Dobrin{\^c}i{\'c} et al.\ 2008 and the reviews of Balick \& Frank 2002 and Kwok
2008).  In addition, recent deep searches for faint PNe have shown that the
fraction having spherical morphologies is higher among low-surface-brightness
nebulae (e.g., Parker et al.\ 2006; Miszalski et al.\ 2008b; Jacoby et al.\
2010); hence the number of known PNe that do not necessarily require a binary
central system is larger now than before these deep searches. Thus the degree to
which binary-star ejection is a major channel for the production of PNe remains
uncertain and controversial; see the reviews by Bond (2000) and De~Marco
(2009). 


In this paper we discuss the properties of the eclipsing binary central star of
the PN SuWt~2, based on observations collected by us over the past two decades.
This object presents a number of challenges to our understanding, as described
above, of PNe with close-binary nuclei. We have presented several unrefereed
progress reports on SuWt~2 in past years as we gradually learned more about the
system (Bond, Ciardullo, \& Webbink 1996; Bond, Exter, \& Pollacco 2001; Bond et
al.\ 2002; Exter et al.\ 2003; Bond et al.\ 2008); in this journal paper we now
discuss our extensive material on this remarkable object in detail.

\section{SuWt~2}

The PN SuWt\,2 (PN\,G311.0+02.4)  was detected by Schuster \& West (1976) on
deep ESO 1-m Schmidt plates taken during emulsion-sensitization tests. (The
nebula had actually been discovered earlier by Andrews \& Lindsay 1967 on
long-exposure plates taken with the ADH Baker-Schmidt telescope, but to our
knowledge this prior discovery has not been referenced in the literature until
now.)  Schuster \& West described the nebula as an elliptical ring enclosing a
prominent and relatively blue central star. Follow-up image-tube spectrograms
obtained by West (1976) confirmed the emission-line PN nature of the nebula,
with a heliocentric nebular radial velocity of $-40 \pm 9\,\rm km\,s^{-1}$.  
West quotes a private communication from N.~Sanduleak, who classified the
central star B9\,V from a Curtis Schmidt objective-prism plate (below, from
more modern data, we slightly revise the spectral type to early~A).

Narrow-band CCD images of SuWt~2 were presented by Schwarz, Corradi, \& Melnick
(1992). The angular dimensions of the bright ring, based on these images, are
$86\farcs5\times43\farcs4$ (Tylenda et al.\ 2003).  Further deep imaging and
spectroscopy has been obtained by Smith, Bally, \& Walawender (2007) and Jones
et al.\ (2009), and these images show faint bipolar lobes extending along an
axis perpendicular to the plane of the brighter ring. This structure is also
seen in the SuperCOSMOS H$\alpha$ image available online (Parker et al.\ 2005).
Jones et al.\ give a nebular systemic radial velocity of $-25\pm5\,\rm
km\,s^{-1}$, and present very deep images of the nebula.   To illustrate the
extended, low-surface-brightness lobes in a wider field of view than shown by
Jones et al., we present in Fig.~1 a deep image in an H$\alpha$+[\ion{N}{2}]
filter, obtained by H.E.B. on 1995 January~31 with the CTIO 1.5-m
telescope\footnote{A pseudo-color version of this image, made from frames in
[\ion{O}{3}] 5007~\AA\ and H$\alpha$+[\ion{N}{2}], is available at
http://hubblesite.org/newscenter/archive/releases/2008/21/image/a/}.

The elliptical ring exhibited by SuWt\,2 is suggestive of a thin annulus viewed
at a high inclination. Assuming the ring to be perfectly circular, the viewing
inclination implied by the axis ratios measured by Smith et al.\ (2007) is
$i=64^\circ\pm2^\circ$, or is $68^\circ\pm2^\circ$ for the proportions measured
by Jones et al.\ (2009). The bright, hollow, thin, and nearly perfectly
elliptical ring, with considerably fainter bipolar lobes, is an unusual
morphology among PNe. However, a morphologically very similar object is the
northern ring PN WeBo\,1.  Its optical central star is a cool barium giant,
which is photometrically variable with a period of 4.7~d (Bond, Webbink, \&
Pollacco 2003); however, unlike the eclipsing SuWt~2, the variability of
WeBo\,1's nucleus is believed to be due to starspot activity on the rotating 
cool star. Another PN with a similar thin elliptical ring structure is SuWt\,3
(West \& Schuster 1980), but its central star is very faint. In a separate
context, thin nebular rings are known around several hot, massive supergiants,
as well as SN~1987A, and have been discussed extensively (Smith et al.\ 2007 and
references therein).

The B9\,V spectral type of the star (as reported at the time of the
discovery of the PN) at the center of the SuWt\,2 nebula indicated an effective
temperature too cool to ionize the nebula. This, along with the PN's morphology,
suggested that there is likely to be an additional star in the system, smaller
and optically fainter than the optical component, but hot enough to emit
ionizing flux.  These considerations, along with a morphology suggestive of
ejection into an equatorial plane viewed at high inclination, made the
optically bright star in SuWt\,2 a prime candidate for a program of photometric
monitoring of PNNi aimed at discovering close binaries.  

Several measurements by H.E.B. during an observing run in 1990 at the 0.9-m
telescope at the Cerro Tololo Inter-American Observatory (CTIO) disclosed no
light variations. However, a 13-night CCD observing run on the same telescope in
1991 resulted in the discovery of two eclipses, separated by 5~days. Continued
monitoring during subsequent observing runs in the 1990s led to the finding that
the eclipses recur every 2.5 days.  Thus the system contains either a hot and a
cool component (as is typical of the other known eclipsing PNNi), in an orbital
period of 2.5~days with one deep eclipse per orbit, or two stars of similar
temperature, with an orbital period of 5~days and two deep eclipses per orbit.
Due to the lack of any obvious reflection-effect variability outside the
eclipses, indicating that heating effects are insignificant, the latter
possibility seemed more likely. Subsequent high-S/N (signal-to-noise) AAT
echelle spectroscopic observations (see below) in fact revealed a double-lined
system containing two nearly equal-brightness stars, confirming that the period
is indeed close to 5~days and that there are two eclipses of comparable depth
per orbital period.


In this paper we assemble and analyze the photometry and optical and UV
spectroscopy that we have obtained over the past two decades. The result is a
determination of the stellar and binary parameters for the eclipsing system,
along with the raising of a number of intriguing astrophysical puzzles.

In the discussion below we use ``SuWt\,2'' to refer to the PN; for the central
star, which has not yet received a formal variable-star designation, we use 
``NSV\,19992'', which is the star's number in the New Catalogue of Suspected
Variable Stars (Kazarovets, Samus, \& Durlevich 1998).

\section{Photometry}

Because the orbital period of NSV\,19992 is very close to 5~days, with an
eclipse duration of more than 17~hours, accumulation of a light curve covering
all orbital phases has been a slow process, involving several different
telescopes. Most of our photometry is differential with respect to several
nearby comparison stars, but on several photometric nights we also measured
absolute photometry. Stringent checks were carried out to ensure that the
various data sets could be combined onto the same system.

Our photometric observations have all been made with a variety of CCD cameras on
telescopes at CTIO\null. Data were obtained with the 0.9- and 1.5-m reflectors
on 63 nights between 1990 and 2001. In 2003 the small- and medium-size
telescopes at CTIO came under the management of the Small and Medium-Aperture
Research Telescopes System (SMARTS) Consortium, and we used the SMARTS 1.3-m
telescope in 2007 for additional observations on 11 nights, as described below.

A finding chart identifying NSV\,19992 and three nearby comparison stars that we
selected, designated C1, C2, and C3, is given in Fig.~2.  Standard data
reduction---bias subtraction and flat-fielding---has been carried out by us, or
for the SMARTS observations by the automated pipeline at Yale University.

\subsection{Absolute Calibration}

Absolute photometry of NSV\,19992 and the comparison stars was obtained with the
CTIO 0.9- and 1.5-m telescopes and the SMARTS 1.3-m telescope on 10 photometric
nights between 1990 and 2007, with a total of 14 individual $BVI$ observations.
Photometric reductions were performed using the {\sc daophot ii} package
(Stetson 1987, 2000) running under IRAF,\footnote{IRAF is distributed by the
National Optical Astronomy Observatory, which is operated by the Association of
Universities for Research in Astronomy (AURA) under cooperative agreement with
the National Science Foundation.} or with simple aperture photometry. 
Calibration to the standard Johnson-Kron-Cousins $BVI$ system was done using
observations of standard fields from Landolt (1992) and transformation routines
in the IRAF {\sc photcal} package.

The calibrated magnitudes and colors for NSV\,19992 and the three comparison
stars are given in Table~1, along with the stars' coordinates from the USNO
NOMAD  astrometric catalog (Zacharias et al.\ 2004). We saw no evidence for
variability of the comparison stars, nor of the outside-eclipse brightness of
NSV\,19992 itself, above a level of $\sim$0.02~mag during this 17-year interval.

\subsection{Differential Photometry}

CCD differential photometry of NSV\,19992 and the three comparison stars in the
$BVI$ filters was taken with the CTIO and SMARTS telescopes in various runs
between 1990 and 2007.  During the later years, as the eclipse ephemeris became
better known, it became possible to observe several eclipses intensively during
entire nights.  We took advantage of STScI's membership in the SMARTS Consortium
by obtaining individual service observations of NSV\,19992 at critical orbital
phases on 11 nights in 2007 April and June, thus filling in phases that still
had not been covered in the earlier runs, as well as improving the precision of
the orbital ephemeris.

Instrumental magnitudes were obtained using the IRAF {\sc apphot}
aperture-photometry routine and manual aperture photometry using {\sc
imexamine}, and PSF fitting magnitudes were obtained using {\sc daophot ii}
running under IRAF\null. 

As shown in Table~1, the comparison stars C1 and C2 are significantly redder
than NSV\,19992; this introduced offsets into the raw differential photometry
because of variations in the effective wavelengths of the bandpasses with the
different cameras and CCDs used over the years, particularly in the $B$ band. In
our final reductions, we used differential magnitudes with respect only to C3,
which has similar colors to NSV\,19992. Color-dependent corrections were made to
the differential measurements based on the transformation equations derived from
the standard stars, and then the adjusted differential magnitudes were added to
the calibrated values for C3 listed in Table~1, to place the final photometry 
of NSV\,19992 on the standard $BVI$ system. Because of the similarity of colors
between NSV\,19992 and C3, we did not correct for trends with airmass. Complete
tables of our photometry will be lodged with
CDS\footnote{http://cdsweb.u-strasbg.fr/astroWeb/astroweb.html}. 

\subsection{Orbital Ephemeris and Light Curve}
\label{sec:ephem1}

Portions of eclipses of sufficient duration for determining times of minimum 
were observed during long runs between 1997 and 2007. We measured the times of
minimum by fitting a Gaussian to the data points around each eclipse for each
filter, with the fitting errors being used to determine formal uncertainties in
the resulting eclipse times. Then, knowing already that the period was close to
5~days, we assigned cycle counts to each eclipse (two of which are secondary
eclipses), and finally performed a weighted least-squares fit using a linear
ephemeris to solve for the orbital period $P$ and time of primary eclipse $T_0$.
The times of eclipse, cycle counts, residuals, and filters used are given in
Table~\ref{tab:toe}.

The final adopted ephemeris is
\begin{equation}
\label{eq:ephem}
{\rm Pri.\ Min.\ \, (HJD)} = 2450668.5915(10) + 4.9098505(20) \, E \, .
\end{equation}
Using the Lafler-Kinman (1965) algorithm, we confirmed that the adopted period,
within the quoted uncertainty, fits the entirety of our data from 1990 to 2007. 

All the 280 $BVI$ data points were then phased to the ephemeris to produce the
light and color curves shown in Fig.~3. On photometric nights the Poissonian
errors are less than $\pm$0.01~mag, i.e., smaller than the plotting symbols in
the figure. However, several of the eclipses were observed on non-photometric
and even cloudy nights, producing a larger photometric scatter under those
conditions. There are also some small residual systematic differences remaining
between the data from different cameras and telescopes, of an uncertain origin
(possibly flat-fielding or vignetting issues), at a level of about 0.02~mag.
These account, for example, for the apparent small excursions just before the
eclipse ingresses in Fig.~3.

The colors are essentially constant throughout both eclipses, and the eclipses
have nearly identical depths; both facts are consistent with the two stars
having similar effective temperatures. The primary eclipse, at orbital phase 0,
is actually very slightly deeper than the secondary eclipse, and as we will show
below the primary star eclipsed at phase 0 is slightly hotter and slightly more
massive than the secondary component, and is a faster rotator. We marginally
detect the light variations due to the slightly ellipsoidal shapes of the stars,
giving rise to small increases in brightness around phases 0.25 and 0.75. The
secondary eclipse occurs at phase 0.5, so there is no evidence for a non-zero
orbital eccentricity.

\section{Stellar Spectroscopy}

Here we discuss a range of spectroscopic observations of NSV\,19992: in the
ultraviolet using the {\it International Ultraviolet Explorer\/} ({\it IUE\/}),
for the purpose of searching for a hot companion of the optical central binary
(of two nearly identical A~stars) and for obtaining the UV spectral energy
distribution (SED); ground-based spectroscopy for the purpose of measuring
radial velocities of the binary; and ground-based spectroscopy for the purpose
of a spectral-type analysis and determination of the optical SED\null. The
reduction of all data followed the standard scheme (bias subtraction,
flat-fielding, sky subtraction), followed by conversion of the spectra to a
constant velocity grid, correction to the heliocentric standard, and a continuum
normalization from a low-order polynomial fit. We used the IRAF and 
Starlink\footnote{http://starlink.jach.hawaii.edu} environments for the data
reduction, and {\sc molly} and {\sc pamela}\footnote{{\sc molly} and {\sc
pamela} are software written by Tom Marsh, Warwick, UK} were used for the
wavelength calibrations and velocity normalizations. Arc spectra generally
bracketed the targeted pointings. Details of the observations and data reduction
are given in Table~3, and below we describe only those details unique to each
run.

\subsection{\it International Ultraviolet Explorer}

As part of a program of UV spectroscopy of optically bright PNNi that appear to
be too cool to ionize their nebulae, M.~Meakes and H.E.B. observed SuWt~2 with
{\it IUE\/} in 1990 and 1991. The first short-wavelength low-dispersion spectrum
was expected to reveal the UV spectrum of a hot PN central star component;
however, the spectrum showed only a faint short-wavelength continuum, which is
compatible with that expected from the A-type components of  NSV\,19992. The
1991 short- and long-wavelength spectra again only showed the continuum of
NSV\,19992, whose flux drops to a nearly undetectable level below
$\sim$1500~\AA\null.  The {\it IUE\/} results were described briefly by Bond et
al.\ (2002).  We return to the {\it IUE\/} spectra below in \S9.4.

In retrospect, we see that the {\it IUE\/} observations were made at orbital
phases of 0.94, 0.68, and 0.70, respectively; thus they were all taken outside
the eclipses of the pair of A~stars.

\subsection{SAAO}

NSV\,19992 was observed in 1995 (by  D.L.P. and F.~Marang) with the SAAO 1.9-m
and the RPCS (Reticon Photon Counting System).  The RPCS was an instrument that
produced two $1\times 1800$-pixel spectra: one of NSV\,19992 and one of the sky.
No nebular lines were visible on the target or sky spectra.  The continuum
signal-to-noise (S/N) ratio for these spectra is around 28. Analysis of the
velocity standard stars observed on the same nights shows there are no drifts in
the radial-velocity scale over the observing run, with a scatter about measured
versus true radial velocity of $\pm$$6\,\rm km\,s^{-1}$.  

\subsection{NTT}

NSV\,19992 was observed in 1995 with EMMI on the 3.5-m ESO New Technology
Telescope (NTT) at La Silla under remote control from Garching, Germany.  We
used a slit $1''$ wide to ensure radial-velocity  accuracy. The spectra were
``optimally extracted,'' i.e., weighted to give the maximum S/N ratio, of up to
100 in the continuum.  No nebular lines were visible on the CCD images. No
radial-velocity standard stars were observed during this run. 

\subsection{AAT}

NSV\,19992 was observed with the UCLES echelle spectrograph on the 3.9-m AAT in 
2000. These nights suffered from scattered clouds and had an average seeing of
$1\farcs8$. The continuum S/N for these spectra is around 60. No radial-velocity
standards were observed during the run.   

These high-resolution AAT echelle spectra are shown in Fig.\ 4; these firmly
confirmed for the first time that the spectrum is double-lined, consisting of
two A-type stars, and thus  established that the 4.9-day period was the correct
one.

Additional data were taken with the RGO spectrograph of the AAT in 2001. The
slit width was set to match the seeing. The continuum S/N ratio for these
spectra is around 50.  Analysis of radial-velocity standards observed during the
run resulted in nightly corrections of $<$$2\,\rm km\,s^{-1}$.  

\subsection{CTIO and SMARTS}

Spectra of NSV\,19992 were obtained by H.E.B. with the Ritchey-Chretien (RC) 
spectrograph on the CTIO 1.5-m in 2001 and 2003 (at which time the telescope had
come under management by the SMARTS consortium), and by SMARTS service observers
in 2006 and 2007. There were no changes in the instrumentation over the
2001-2007 interval. The slit width was $2''$ for the former two years and $3''$
for the latter two.  Nebular contamination of the  stellar spectra at the Balmer
lines was obvious for these data, and was removed using the spectra from the
nebular regions close to the star on the CCD\null.  Flux-calibration standards
(Hiltner\,60 and LTT\,4364) were observed for all these observations.  The
reduction of these data hence included  flux calibration, with a scatter in the
sensitivity function of $\pm$0.01\,mag. Note that slit losses were not corrected
for, and we are uncertain if the conditions were truly photometric. Therefore,
we cannot guarantee the absolute flux scale is correct.

\section{Spectral Analysis}
\label{sec:spt}

\subsection{Spectral Energy Distribution}
\label{sec:fluxdist}

We have compiled the SED of NSV\,19992 from the UV to the NIR  by
combining data from several sources. The UV fluxes were obtained from the
low-resolution {\it IUE\/} spectra described in \S4.1.  The spectra were binned
into 50~\AA\ regions, and only wavelength ranges with S/N $>$ 10 were retained.
In the optical region we used the broad-band magnitudes given in Table~1 
and the flux-calibrated spectra (\S4.5). For the NIR  we used
broad-band photometry from the DENIS and 2MASS surveys (listed in Table~1).

We then fit reddened Kurucz (1993) model flux distributions to the SED\null. 
The flux distribution is well fitted using a single-star model, as expected
because of the close similarity of the two stars implied by the light curve. 
The best fit was obtained with an effective temperature of $\teff =
9500\pm500$~K, a surface gravity of $\log g = 4.0\pm0.5$, and reddening of
$E(B-V) = 0.40\pm0.05$.  The combined energy distribution is shown in
Fig.\,\ref{flux_plot}, along with the best-fitting Kurucz model. Direct
integration of the observed flux distribution gives $\ftot = (1.25 \pm 0.20)
\times10^{-9}$ ergs\,s$^{-1}$\,cm$^{-2}$, corrected for $E(B-V) = 0.40$.  The
Infrared Flux Method (Blackwell \& Shallis 1977) gives a similar result of
$\teff = 9430 \pm 560$\,K and a combined angular diameter of the two stars of
$\theta = 0.022 \pm 0.003$\,mas.  

The corresponding spectral types for both stars would be near A1~V\null. The
dereddened color of the combined light is $(B-V)_0=0.02$, consistent with the
spectral type.

\subsection{Spectral Diagnostics and Rotational Velocities}

As described below in \S\ref{sec:echelle}, from the AAT echelle spectra we
disentangled the spectral lines from the two components of the binary into
spectra of the individual stars. From the individual spectra we measured
equivalent widths of several unblended lines in the 4460--4500~\AA\ range. These
were used with the LTE spectrum-synthesis code UCLSYN (Smith 1992; Smalley,
Smith, \& Dworetsky 2001) to determine stellar parameters and chemical
abundances.

The parameter space around that indicated by the SED analysis described above
was searched to find the best self-consistent solution for (i)~the
\ion{Fe}{2}/\ion{Fe}{1} ionization balance, (ii)~the microturbulent velocity
(\mictrb) from the \ion{Ti}{2} lines, and (iii)~a null-dependence of the results
on excitation potential.  The resulting stellar parameters for the components of
NSV\,19992 and their chemical abundances are given in Table~4.  The star we have
designated as the primary is marginally hotter than the secondary, but the
difference is not statistically significant. 

Plots of the model fits to the \ion{Mg}{2} 4481~\AA\ and  H$\alpha$ lines of the
echelle spectra are shown in Fig.\,\ref{f:mg_plot}. 

We fitted the \vsini\ of both stars while running UCLSYN, assuming an
instrumental FWHM of 0.14~\AA\null. The lines of the primary star are clearly
rotationally broadened, with $v \sin i = 17\,\kms$; in the discussion below we
sometimes refer to the primary as the ``fast'' rotator. For the secondary star
we only obtain an upper limit to $v \sin i$ of $5\,\kms$. The derived value of
\vsini\ is not strongly dependent on the adopted physical parameters. The
luminosity ratio was calculated at 4480\,\AA, and is not significantly different
from equal luminosities of the two stars.

There is a hint that the \ion{He}{1} 4471\,\AA\ line is  present in the slow
secondary, but not in the fast primary.  However, the S/N ratio is rather poor
for this faint feature and the identification of \ion{He}{1} in the secondary
star is uncertain. The \ion{Mg}{2} 4481~\AA\ line was difficult to fit with a 
solar abundance, but a value of  $\rm[Mg/H] \simeq +0.4$~dex gives a good fit to
the observed line profiles. The Mg II 4481 line is known to exhibit marked
deviations from LTE (Przybilla et al. 2001), which could account for the
apparent overabundance. A value of [Ti/H] of  $+0.59\pm{0.15}$ was formally
derived for the primary star. However, we consider this finding to be more
uncertain than the errors reflect; NLTE effects may be important.

\section{Spectroscopic Orbit}
\label{sec:spec}

From our high-resolution (UCLES, RGO, RPCS) spectra we measured the radial
velocities of the two stars in the binary, leading to a determination of the
orbital parameters and the stellar masses. There was some interplay between the
spectroscopic and photometric analyses, allowing a narrowing of the range of
possibilities based on the other's results. 

\subsection{Measuring the Radial Velocities}

The AAT echelle UCLES spectra were the most important for fitting the radial
velocities with a binary orbit model, as they are at excellent velocity
resolution. Their phase coverage, however, is a little sparse, and this is where
the AAT RGO mid-resolution spectra  become useful. The SAAO RPCS
lower-resolution spectra were also useful, despite the low velocity precision
and lower SNR, because they have good orbital coverage; it is also the case that
as they were taken 5 years before the echelle spectra they could be used to
check for any drift in the systemic velocity of NSV\,19992 (for example, as
expected if a third star were present in the system). To this particular study
we have added three low-resolution spectra taken with the NTT.

All velocities were corrected to the heliocentric frame. The AAT mid-resolution
and SAAO data have been checked for velocity offsets using observations of
radial-velocity standards.  In Table~5 we list the velocities, and next we
describe how they were measured. 

\subsubsection{Cross-Correlation for the AAT Echelle Spectra} 
\label{sec:echelle}

Radial velocities were determined from the high-resolution echelle spectra using
cross-correlation with a template spectrum. The template spectrum was created
from those which we modeled for each star of the binary. The main difference
between the two stars is their rotational velocities, and this was accounted for
in the templates. 

The cross-correlation functions (CCFs) show two very distinct peaks, which are
clearly resolved in all but one case.  To measure the position of the peaks in
the CCFs and check that the rotational broadening that had been applied to the
spectra was sufficient, we calculated two model CCFs of the synthetic spectrum
(one for each star)  against two versions of itself after convolution with the
rotational broadening function of Uns\"{o}ld (1955). The broadening necessary
for each star are consistent with the values found already: 15\,km\,s$^{-1}$ for
the fast rotator and none for the slow rotator.  We then used a least-squares
method to optimize the fit to the CCFs using the sum of the two model CCFs. The
free parameters for the fit are the radial-velocity shift between the model CCF
and the star and a scaling factor for each CCF\null. An example of one such fit
is shown in Fig.\,7. Also shown are the echelle spectra from each component
after being ``disentangled'' (Hynes \& Maxted 1998), revealing each component's
spectrum alone.

\subsubsection{Gaussian Line Fitting}

Gaussian fitting of individual absorption lines was used to measure the radial
velocities from the lower-resolution  AAT and SAAO spectra.  For this we
focussed on the \ion{Mg}{2} 4481~\AA\ line and did the fitting in the {\sc
molly} environment.  We determined the wavelength of the \ion{Mg}{2} feature
from the theoretical spectra for each star (\S\ref{sec:spt}), broadened and
resampled to match the AAT and SAAO spectra ($\lambda_0=4481.2195$\,\AA\ in all
cases).  We also fit to the Balmer lines at the phases around quadrature, where
the lines from each star somewhat separate on the observed spectra, but only as
a check on the superior \ion{Mg}{2} results.
  
The theoretical spectra used in the cross-correlation (the echelle spectra) are
of zero velocity, so for these the systemic velocity (that of the whole binary
with respect to us) should be on the correct scale.   To ensure that the
Gaussian-fit velocities (the mid-resolution spectra)  were on the same scale we
compared the cross-correlation results for the echelle spectra with Gaussian-fit
results for the echelle spectra, from which we found a value 3.3\,km\,s$^{-1}$
to be  subtracted from the Gaussian-fit velocities.

\subsection{Orbit Solution}

We used the double-lined spectroscopic binary orbital-solution code {\sc
binary},\footnote{available at
\url{http://www.chara.gsu.edu/$^{\scriptscriptstyle\sim}$gudehus/binary.html}} 
developed by Gudehus (2001), to fit for the orbital parameters of NSV\,19992. We
fitted each observing run of radial-velocity measurements separately in order to
account for possible shifts in the center-of-mass velocity of the system with
time. These possible shifts are discussed again in \S 7.3.

We began by fitting the radial velocities determined from the AAT echelle data
obtained in 2000, as these are the highest-resolution spectra and thus yield the
highest-precision radial-velocity measurements. We held the orbital period fixed
to the already-determined ephemeris (Eq.~1), but fit for all other orbital
parameters.  This fit formally yields a slight, and marginally significant,
non-zero eccentricity of $e=0.023 \pm 0.011$.  However, as noted at the end of
\S 3.3, the light-curve eclipse times show that the secondary eclipses occur at
phase 0.5 (see also the $O-C$  residuals of the eclipse timings in Table
\ref{tab:toe}), giving no support for a non-zero eccentricity. Moreover, the
eccentric orbit solution predicts eclipse times that are inconsistent with the
observed eclipse times.  From our data we cannot fix more accurate limits for
the eccentricity; hence for this paper we fix it to a value of 0.0.  We refit
the AAT echelle radial-velocity data with $e=0$ and obtained the orbital
parameters listed in Table~6. A follow-up study of the small but possibly
non-zero eccentricity could add insight into the history of this system. 

Next, we fit each of the other three epochs of radial-velocity data (two runs in
1995, one in 2001) by fixing all orbital parameters to  the values determined
from the above fit, save for the center-of-mass velocity, $v_\gamma$. We did
this to look for any systemic velocity offset of each epoch relative to the AAT
echelle epoch. The resulting best-fit systemic velocities are tabulated in
Table~7, where indeed we see evidence for systemic velocity shifts of up to
$\sim 40$ km s$^{-1}$. We discuss these systemic velocity shifts in the context
of a possible third body in the NSV\,19992 system below (\S\ref{sec:3rdbody}).

We subtracted the best-fit $v_\gamma$ from each epoch and performed a final fit
to all of the radial-velocity data together. The resulting orbit parameters are
summarized in Table~8 and the fit is displayed in Fig.~\ref{fig:rvfitfin}. 

The components of the NSV\,19992 eclipsing binary evidently have nearly
identical masses. The formally more massive primary ($M_1$) is identified as the
rapid rotator (i.e., the ``fast" star). We emphasize, however, that the best-fit
mass ratio of $q = M_2/M_1 = 0.9922 \pm 0.0044$ (Table~9) is consistent with
unity at the $\approx 2\sigma$ level and thus we cannot definitively establish
which component is truly the more massive one. As we discuss in
\S\ref{sec:evol}, the rapid evolutionary timescales of the components make it
probable that the true mass ratio is nearly exactly 1.0.

\subsection{Evidence for a Third Body\label{sec:3rdbody}}

As noted above, we observe changes in the systemic velocity from one epoch to
the next: $-43\,\rm km\,s^{-1}$ in 1995.3, $-21\,\rm km\,s^{-1}$ in 1995.4,
$-8\,\rm km\,s^{-1}$ in 2000.4, and $-4\,\rm km\,s^{-1}$ in 2001.2. The first
value is based on an observational study of only three spectra with no
radial-velocity standards also observed. Hence its uncertainty is larger than
the formal error. We address this in the discussion that follows.

These different systemic velocity values suggest that the NSV\,19992 eclipsing
binary (EB) itself orbits a third body. With only four distinct epochs it is
impossible to determine the period or amplitude of these variations. However,
the rapid change observed between the 1995.3 NTT and 1995.4 SAAO data, separated
by only 0.10~yr, suggests a relatively short-period orbit, of order a few
months. If we assume that the observed range of $v_\gamma$ values of about
$40\,\kms$ represents the extremes of the EB's orbit about the putative third
body, then the mean systemic velocity of NSV\,19992 is near $-23 \, \kms$
(assuming a circular orbit), consistent with the nebular radial velocity of
$-25\pm5\,\kms$ (Jones et al.\ 2009). 

If we further assume that the time between the NTT and SAAO data epochs (1995.3
and 1995.4)  represents $\sim$1/2 of the  orbital period about the third body,
this orbit can be characterized by a period of $P_3 \approx 0.2$~yr and a
velocity semi-amplitude of $K_{\rm EB} \approx 20\, \kms$. The mass function for
the third body is then given by \begin{equation} f(M) = {M_3^3 \, \sin^3 i_3
\over (M_3 + M_{\rm EB})^2} = 0.061 \, \Biggl({K_{\rm EB} \over
20\,\kms}\Biggr)^3 \,  \Biggl({P_3 \over 0.2\,\rm yr}\Biggr) \, M_\odot\, ,
\end{equation} where $M_3$ is the mass of the third body, $M_{\rm EB}=M_1+M_2$
is the total mass of the EB ($\sim$$5.4\,M_\odot$ according to the previous
subsection), and $i_3$ is the inclination of the third body's orbit, assumed to
be circular. In the evolutionary scenario that we outline in \S9.5 the system is
a triple in which the initially more massive star has formed the PN and so it
should be orbiting in the plane of the nebular ring, which is viewed at an
inclination of $\sim$$68^\circ$. Thus we expect $\sin i_3 \approx 0.9$.

To yield the above $f(M)$ therefore requires an unseen mass 
$\sim$$1.54\,M_\odot$\null. Even if the orbit is viewed edge-on, the mass is
about $1.41\,M_\odot$\null. Our result is problematic, not only because it is at
or above the Chandrasekhar limit,  but also because the evolutionary scenario 
proposed below (\S9.5)---in which the third body was originally a slightly more 
massive star of $\sim 2.9\,M_\odot$---requires a considerably lower remnant
mass, of about $0.7\,M_\odot$.

The discrepancy would be alleviated if we gave low weight to the outlying
systemic velocity from the low-resolution NTT spectra of 1995; as we noted in
\S4.3, these had not been accompanied by observations of velocity standard
stars, and there were only three observations. Without that data point, we would
have no evidence for $K_{\rm EB}$ being larger than $\sim$$10\,\kms$, and a much
lower remnant mass would be compatible with $f(M)$, even if the orbital period
is significantly longer than 0.2~yr.   Indeed, taking $K_{\rm EB}=10\,\kms$ and
holding the other terms in Equation (2) the same as above, gives $M_3 \approx
0.73\,M_\odot$. This estimate becomes $M_3 \approx 0.85\,M_\odot$ if
$P_3=0.3$~yr.

\section{Nebular Reddening}

A detailed study of the kinematics of the nebula has been carried out by Jones
et al.\ (2009).  However they did not measure the nebular reddening, which we 
wish to compare to the stellar reddening of NSV\,19992. The aim here is to see
how similar they are---very different values would be an argument for the PN and
the binary being a chance line-of-sight alignment. 


We used an observation from the AAT's RGO 25cm+TEK 250B grating, taken on 1995
January~7. The  spectra were taken at a PA of $0^\circ$ and with an exposure
time of 1200\,s. Arc spectra (Cu-Ar) were taken as well as wide-slit spectra of
the flux-standard star BD$+8^\circ$\,2015. The standard  reductions were
followed, including a 2D wavelength calibration. The wavelength coverage was
3760--7394 at 3.6\,\AA\,pix$^{-1}$. Flux calibration of the wide-slit SuWt\,2
spectral image was performed and transferred to the narrow-slit spectral image.
Then a  very narrow region of the 2D narrow-slit spectrum of SuWt\,2 was
extracted, close to the central star and thus within the nebular ring.   Sky
subtraction for this extracted spectrum could not be done from these
observations because the telluric lines were contaminated  by nebular emission.
Instead we used a scaled  spectrum of the telluric emission taken from an
observation of another star on the same night.   No extinction or airmass
corrections were applied. 


To calculate the nebular reddening, $c(\rm H\beta)$, we compared the measured
value of H$\alpha$/H$\beta$ to the predicted ratio for nebular gas of 10000\,K
and $\log(n_e)=4$ (Hummer \& Storey 1987). Contamination of the H\,{\sc i} lines
by He\,{\sc ii} for this nebula is less than a few percent and a similar
percentage for H$\alpha$ and H$\beta$.  
We adopted the Galactic extinction law of Howarth (1983).  The Balmer ratio,
$5.18\pm{0.78}$, results in a value of $c(\rm H\beta)$ of $0.83\pm{0.20}$. 

To compare this value to the stellar reddening we can use the conversion
$E(B-V)=c(\rm H\beta)/1.492$, which is based on the Howarth reddening law,
yielding $E(B-V)=0.56\pm{0.14}$. This is formally larger than the stellar value
of $E(B-V)=0.40\pm0.05$, but consistent at just outside the 1$\sigma$ level. One
should also bear in mind that when comparing the stellar and nebular values we
make a number of approximations: (i)~we assume that all the nebular reddening is
interstellar; there will be some internal reddening although it is normally
considered to be fairly low compared to the interstellar, (ii)~the conversion
between $c(\rm H\beta)$ and $E(B-V)$ can vary, as it depends on the value of $R$
(total-to-selective extinction) and the stellar energy distribution. As an
example, Kaler \& Lutz (1985) calculated conversion constants of 1.43 to 1.68;
the latter would bring the nebular $E(B-V)$ down to 0.49.

\section{Light-Curve Solution}
\label{sec:ephem}

The photometry and  light curve of the eclipsing system provides information on
the orbital inclination, relative radii, relative temperatures, and apparent
magnitudes of the two stars and their reddening.  In this section we explain the
procedures used to  determine final stellar and orbital parameters for
NSV\,19992 from our photometry.

We performed a simultaneous fit to the $BVI$ light curve data and the
radial-velocity data, using modeling tools described in detail in several recent
EB analyses (e.g., Stassun et al.\ 2004, 2006, 2007, 2008;  G\'{o}mez Maqueo
Chew et al.\ 2009).

Briefly, we used the most recent version of the EB light-curve modeling code of
Wilson \& Devinney (1971, updated 2007), as implemented in the {\sc phoebe}
package of Pr\v{s}a \& Zwitter (2005).  We adopted the ephemeris for the system
derived in \S\ref{sec:ephem1} and held it fixed throughout the fitting
procedure. We additionally held the orbital eccentricity fixed at zero; adopting
the slight non-zero eccentricity found from fitting the AAT radial-velocity data
alone, $e=0.026$ (see \S 6.2), changes the results presented here negligibly
relative to the uncertainties.

In order to maintain control of the solution and its many free parameters, we
performed this fitting in stages. Initially we held the orbital parameters fixed
at the values determined from the double-lined spectroscopic orbit
(\S\ref{sec:spec}), and fixed the effective temperature of the primary component
to the value determined from our spectral analysis (9250~K, \S\ref{sec:spt}).
This allowed us to determine an initial solution for the temperature of the
secondary component and the sum of the component radii. Next we added the system
inclination as a fitted parameter. We assumed a linear cosine limb-darkening
law, with coefficients linearly interpolated from the tables of Van~Hamme
(1993). In addition we assumed an albedo of 1.0 and a gravity-brightening
coefficient of 1.0 for both components, as appropriate for hot stars. We held
the rotation rates of the two stars at their spectroscopically  determined
values of 17 and $<5\,\kms$, respectively. Finally, we performed a fit of all
free parameters together, with the limb-darkening coefficients adjusted between
iterations. 

Due to  the lack of a strong reflection or geometric effect in the light curves,
and considering our assumed circular orbit, we are unable to constrain the
individual component radii from the light  curves alone, only their sum.  Thus,
in our light-curve fitting we solved for the sum of the radii, and then
separately determined the ratio of the radii by requiring the ratios of the
radii  and the temperatures together to yield the correct luminosity ratio as
determined from the spectra (see \S\ref{sec:spt}).

The final light-curve fits are shown in Fig.~\ref{fig:lcfit} and the resulting
component parameters are summarized in Table~9. We see that the eclipsing system
consists of two nearly identical early A-type stars, both with masses near
$2.7\,M_\odot$, and nearly identical temperatures and radii which place them
just above the main sequence. 

The combined radius of the two stars calculated here is $8.2\,R\sun$ (Table~9),
which compares fairly well to the value $5.4\,R\sun$ obtained from the angular
diameter calculated in \S5.1 (0.022\,mas) and distance that will be calculated
in \S9.2 (2.3\,kpc), combined with the small-angle approximation. The
radiometric radius in \S5.1 is of course very sensitive to the adopted effective
temperature.

\section{Discussion}

\subsection{Evolutionary Status of the Components of NSV\,19992
\label{sec:evol}}

In Fig.~\ref{fig:radteff} we compare the measured temperatures and radii of the
two components of NSV\,19992 with Yonsei-Yale (``Y$^2$") evolutionary tracks
(see Demarque et al.\ 2004 and references therein).  The upper panel shows the
early evolution  of a single star of mass $2.69\,M_\odot$ (the mean mass of the
nearly identical components) and solar metallicity, starting from the zero-age
MS (lower left corner), across the Hertzsprung gap, and to the base of the
red-giant branch. The dotted curves represent the same evolutionary track but
for metallicities of $\rm[Fe/H] \pm 0.14$, representative of the $1\sigma$
uncertainty in the measured Fe abundance of the system (see Table~4). The two
overlapping filled circles show the effective temperatures and radii of the
components of NSV\,19992. Gratifyingly, both points lie precisely on the track
corresponding to their dynamical masses.


Note that the very small errors on the temperatures represent the highly
accurate temperature {\it ratio\/}  determined from the light-curve solution;
however, there is in addition a systematic error of 250~K on the absolute
temperature scale (see Table~4), which  is indicated in the panel. As the
individual radii are connected by the accurately determined radius sum and ratio
(see Table~9), their  {\it relative\/} placement in the figure cannot be
changed by more than the error bars shown on  the individual points.  

For visual simplicity, we have plotted the evolutionary tracks for only a single
mass (that of the mean mass of the components); the difference in the
positioning of the tracks for masses within the $1\sigma$ mass uncertainties is
less than that shown for the $1\sigma$ range in [Fe/H] (dotted tracks). Finally,
for reference, absolute ages (in Myr) from the model are shown at several points
along the track. The implied ages of the two components are close to 520~Myr.

Both components of NSV\,19992 appear to occupy a remarkably short-lived phase of
evolution. Even with the systematic uncertainty on the absolute temperatures of
the components, it is clear that they reside in the ``blue hook" that occurs
just prior to the Hertzsprung gap, and which lasts in total only $\sim$1~Myr. In
spite of the brevity of this evolutionary phase, a small number of other EBs are
also known in a similar state of evolution, including SZ~Cen (Popper 1980),
V1031~Ori (Andersen et al.\ 1990), and WX~Cep (Popper 1987). As discussed by
Andersen et al.\ (1990), it is surprising that so many systems have been seen in
such a very short-lived state of evolution. One possible explanation is that the
interior physics of the theoretical models (e.g., convective overshooting) could
move the stars to be on the ``red hook" instead of the ``blue hook," the former
being a somewhat longer-lived stage ($\sim$10~Myr instead of $\sim$1~Myr). 

The evolutionary rapidity in this portion of the H-R diagram can be used to
further constrain the mass ratio of the components. By examining the range of
stellar masses (at solar metallicity) for which the Y$^2$ models predict
temperatures and radii consistent with the measured values, we find that a
minimum mass ratio of $q = 0.9996$ (at $1\sigma$ confidence) is required to
place the two stars this closely in temperatures and radii.

We point out that these stellar evolutionary models represent the evolution of
single stars only. The evolution through the blue-hook phase depends heavily on
the details of convection and mixing which are still treated somewhat
simplistically in the models. For binary systems, with an (as-yet) unknown
mass-transfer history, the situation will be even more complex.  However, a
detailed examination of constraints on the model physics is beyond the scope of
this observational paper.

\subsection{Distance, Nebular Size and Age, and Source of Ionization}

Using the measured temperatures and radii for the components of NSV\,19992 we
can estimate the distance to the system. We find a bolometric absolute magnitude
for the combined light of both stars of $M_{\rm bol} = -1.09$, which, together
with a bolometric correction of ${\rm BC}_V = -0.07$, gives $M_V = -1.02$. The
observed $V=12.00$ and $E(B-V)=0.40$ yield an unreddened $V_0 = 10.76$, from
which we derive a distance $d=2.3 \pm 0.2$ kpc (including uncertainties in the
stellar radii and temperatures, and in the reddening). 

At this distance, the measured semimajor axis of the SuWt~2 nebula of $43''$
corresponds to a physical ring radius of 0.47~pc. Jones et al.\ (2009) have
found a nebular expansion velocity of $28\,\kms$, which when combined with the
radius implies a dynamical age for the ring of $\sim$17,000~yr. 

Interestingly, Smith et al.\ (2007) have speculated that the nebula may in fact
be externally ionized by UV light from the angularly nearby 7th-mag B supergiant
HD~121228 (the very bright star seen in Fig.~1), rather than from the central
stellar system.  HD~121228 has been classified B2~Ib by Humphreys (1975) and
B1~II by Garrison et al.\ (1977). Of course, an early B supergiant could not be
coeval with A-type stars having ages of 520~Myr, so the association would have
to be one that occurred by chance.  The $73''$ angular separation of HD~121228
and the central star of SuWt~2 corresponds to a projected physical separation of
only 0.8~pc, if the two are at the same distance.  The distance to HD~121228 has
been estimated to be 2.8~kpc by Westin (1985), based on Str\"omgren photometry.
Although somewhat larger than the distance we derive for SuWt~2, it is probably
not inconsistent, within the respective uncertainties. However, combined with
the requirement for a chance approach to the nebula by a much younger star, and
the fact that an external source of ionization still leaves the source of the
nebular ejection unexplained, we remain doubtful of a physical association of
the nebula and HD\,121228.  We prefer an interpretation in which the pair of A
stars has a companion which both ejected and ionized the nebula, as described in
more detail below.

\subsection{Non-synchronous Rotations of the Components}

With an orbital period of 4.9~days, the components of NSV\,19992 might be
expected to be rotating synchronously with the orbit. Indeed, all of the A-type
binaries with orbital periods shorter than 3~days in the study of Matthews \&
Mathieu (1993) were found to be circularized and synchronized, as well as most
binaries with periods shorter than $\sim$10~days. However, our study has
demonstrated that, while the orbit of NSV\,19992 is very likely  circular, both
components rotate much more slowly ($\sim$17 and $<$$5\,\kms$ for the primary
and secondary, respectively) than the synchronous rotation speed of $\sim$$43 \,
\kms$.

The system is similar in this respect to V1031~Ori, a 3.4-day eclipsing system
which is one of the very few close binaries in which the stars are known to
rotate more slowly than synchronously (Andersen et al.\ 1990). A possible
explanation is that the stars were once in synchronous rotation, but have slowed
down through conservation of angular momentum as their radii rapidly expanded
upon leaving the MS\null. However this by itself does not explain the dissimilar
rotation rates of the two components in NSV\,19992---considering that they are
of nearly identical mass and radius---unless the change in internal structure of
these stars is an extremely strong function of mass and age.

We note that the primary (``fast'') component of NSV\,19992 has a comparable $v
\sin i$ to what would be expected for synchronous rotation on the MS, when it
had a radius of $\sim$$1.7 \, R_\odot$\null.  Thus we may speculate that the
system may originally have been a wider, eccentric system, in which only the
primary component was pseudo-synchronized (i.e., synchronized to its orbital
angular velocity at periastron passage), and that the orbit only very recently
became circularized when the NSV\,19992 binary became embedded within the
red-giant envelope of the tertiary companion.

In any event, NSV\,19992 offers the potential for a detailed examination and
stringent test of tidal evolution and CE evolution theory. 

\subsection{The Case of the Missing UV Remnant}

As described in \S4.1, at the time of our {\it IUE\/} spectroscopy at the
beginning of this project we were surprised by the lack of a detected hot
companion in the binary. Here we re-examine whether this surprise was justified,
in the light of what we have learned about the object. 

In Fig.~11 we show the short-wavelength {\it IUE\/} spectrum of the SuWt~2
nucleus (from image SWP~41915), smoothed into 10-\AA\ bins and plotted as a
histogram. For comparison, we obtained from the {\it IUE\/}
archive\footnote{available at http://archive.stsci.edu/iue/} the
short-wavelength spectra of three nearby PNNi with well-determined distances:
NGC~246 ($d=495$~pc; Bond \& Ciardullo 1999), and NGC~6853 and NGC~7293 ($d=405$
and 216~pc, respectively; Benedict et al.\ 2009). The respective image numbers
are SWP~42226, SWP~40066, and SWP~42066. We scaled these spectra to the 2300~pc
distance and $E(B-V)=0.40$ reddening of NSV\,19992,  using the formulae of
Cardelli, Clayton, \& Mathis (1989).  These scaled and reddened spectra are also
plotted in Fig.~11.

The figure shows that a relatively hot and luminous PNN like that of NGC~246
would have been detected easily if it were a companion of NSV\,19992.  However,
the more evolved and fainter PNNi of NGC~6853 probably, and NGC~7293 certainly,
would not have been detected. Thus the lack of a detection of a central star in
the UV here is not so unusual;  the existing UV spectra do not rule  out its
presence.

\subsection{An Evolutionary Scenario}


The principal result of our study is that the optical central star of SuWt~2
consists of a close pair of A-type stars, both with masses close to
$2.7\,M_\odot$\null. Both of them have recently left the MS, and the system's
age is about 520~Myr. Thus, neither star is a stellar remnant, and neither one
could plausibly have ejected the surrounding PN, nor could either have supplied
the UV flux required to ionize the nebula. A scenario in which one of the
A~stars is actually a stellar remnant in the process of a ``born-again''
thermal-pulse event also appears to be excluded.

We therefore need a third star, whose initial mass was greater than
$2.7\,M_\odot$, which has evolved to the AGB stage, ejected the PN, and is now
an optically faint PNN or WD\null. Since we have presented evidence that the
center-of-mass velocity of the central binary varies with time, we suggest that
this unseen third body is in fact the remnant core of the star that ejected the
nebula. The likely period of this third body is short enough that  the binary
could have  been inside the envelope when it was an AGB star: a common
envelope.     The system, under our scenario, was thus a hierarchical triple.
For the third star---the original primary---to have completed its evolution to
the PN stage in $\sim520$~Myr, its initial mass need have been only slightly
higher than the masses of the A-type pair, of order $\sim2.9\,M_\odot$\null. In
the case of single-star evolution, such a star would produce a WD remnant of
$\sim0.7\,M_\odot$ (e.g., Kalirai et al.\ 2008, Fig.~12), but the mass could be
lower if the evolutionary growth of the core mass was truncated through a CE
interaction. 

Our picture is that the initial triple was wide enough for the primary to be
able to evolve to the AGB stage, but then it engulfed the pair of A stars into a
CE\null. Due to the high combined mass of the close A~pair, the interaction must
have been led quickly to a catastrophic ejection of the envelope, primarily into
the orbital plane (e.g., as predicted in three-dimensional hydrodynamical CE
simulations---for example Sandquist et al.\ 1998; Morris \& Podsiadlowski 2006;
Nordhaus \& Blackman 2006), producing the very high density contrast between
equator and poles that we observe in the ejected nebula. While most currently
known post-CE PN central stars have short periods, of the order of hours to a
few days, in this case the rapid ejection would have left the post-CE system
still with a fairly long orbital period, as observed---assuming, of course, that
the unseen third body in the SuWt~2 system is indeed the remnant core of the
original primary 

The orbital planes of the close EB and the more distant third body, in our
scenario, need not be coplanar. Thus it is not necessarily surprising that the
orbital inclination of the eclipsing pair, $81\fdg5$, differs from the viewing
angle of the nebula of $\sim$$68^\circ$.

To explain the lack of detection of the remnant in the {\it IUE\/} spectrum, we
note that stellar remnants of $\sim$$0.7\,M_\odot$ cool and fade rapidly (e.g.,
Vassiliadis \& Wood 1994).  Thus it is not implausible that the remnant could
have faded to the UV brightness of the central stars of NGC~6853 or NGC~7293, as
depicted above in Fig.~11, especially given the fairly advanced nebular age of
$\sim$17,000~yr. In this picture, the PN was ionized earlier by UV radiation
from the hot remnant, but is now beginning to recombine. A fast stellar wind
from the hot remnant in its luminous stage would have inflated the faint bipolar
lobes now seen lying perpendicular to the orbital plane.

In summary, our scenario suggests that the SuWt~2 system is the immediate
product of a catastrophic event in which an orbiting  combined mass of more than
$5\,M_\odot$ was ingested into the envelope of an AGB star, leading to ejection
of the envelope into the orbital plane. The distant future of this system may be
destined to be even more catastrophic. As the A stars evolve, they may merge
into a single star, which will then expand to become a red giant or AGB star,
engulfing the third body into a second common envelope, which will likewise be
ejected. The product may be a close pair of white dwarfs whose combined mass
exceeds the Chandrasekhar limit, so that ultimately the system may perish as a
Type~Ia supernova.

\subsection{Future Work}

The most obvious two goals of future work on NSV\,19992/SuWt\,2 are: 

\begin{itemize}

\item To measure the variations of the systemic velocity with time, to confirm
its variability and better constrain the mass of the possible third body in the
system.  Unfortunately, the reflex motion of the eclipsing pair for the unseen
masses discussed above is only of order 1~light-minute, which is smaller than
the typical $\sim$3~min uncertainties in our eclipse timings, so we cannot
verify the presence of a third body from our timing data. Radial-velocity
monitoring will require much patience because of the almost 5-day orbit and the
lack of many facilities for long-term spectral monitoring. 

\item  To obtain UV data to directly detect a third, hot body. As we have
pointed out, the lack of a detection of a central star of SuWt\,2 in the
existing UV from {\it IUE\/} does not provide a strong constraint.
Unfortunately, the low-latitude field surrounding SuWt\,2 cannot be observed by
GALEX, according to the brightness-checking
tool.\footnote{http://sherpa.caltech.edu/gips/tools/chkbstar.html}  However,
deeper UV spectroscopy than has been obtained with {\it IUE}, e.g., with the
{\it HST\/}, might be capable of revealing the hot companion.

\end{itemize}

\acknowledgments

H.E.B. thanks Michael Meakes for making the {\it IUE\/} observations described
here, Robin Ciardullo and Noam Soker for discussions, and numerous telescope
operators, support personnel, and service observers at CTIO, and the SMARTS
telescope schedulers at Yale and Stony Brook Universities\null. He also thanks
the STScI Director's Discretionary Research Fund for partial support of this
work. We thank Tom Marsh for the use of {\sc molly} and {\sc pamela}. We thank
Tanmoy Laskar for his provision of standard-star theoretical spectra, and
Kenneth De Smedt-Van Gent and Joris Vos for  their thorough analysis of the
standard-star radial velocities.   K.G.S.\ acknowledges the support of NSF
Career grant AST-0349075, NSF AST-0849736, a Cottrell Scholar award from the
Research Corporation, and the hospitality of the Space Telescope Science
Institute's Caroline Herschel Distinguished Visiting Scholars program.  Some of
the data presented in this paper were obtained from the Multimission Archive at
the Space Telescope Science Institute (MAST)\null. STScI is operated by the
Association of Universities for Research in Astronomy, Inc., under NASA contract
NAS5-26555. Support for MAST for non-{\it HST\/} data is provided by the NASA
Office of Space Science via grant NAG5-7584 and by other grants and contracts.
This research has made use of the SIMBAD database, operated at CDS,  Strasbourg,
France.
This research used the facilities of the Canadian Astronomy Data Centre operated
by the National Research Council of Canada with the support of the Canadian
Space Agency. This paper uses observations made at the South African
Astronomical Observatory (SAAO).

{\it Facilities:} 
\facility{SMARTS:1.5m},  \facility{SMARTS:1.3m}, \facility{CTIO:0.9m},
\facility{CTIO:1.5m}, \facility{IUE}, \facility{SAAO}, \facility{AAO},
\facility{ESO}


\clearpage


\begin{figure}[h]
\epsscale{0.9}
\plotone{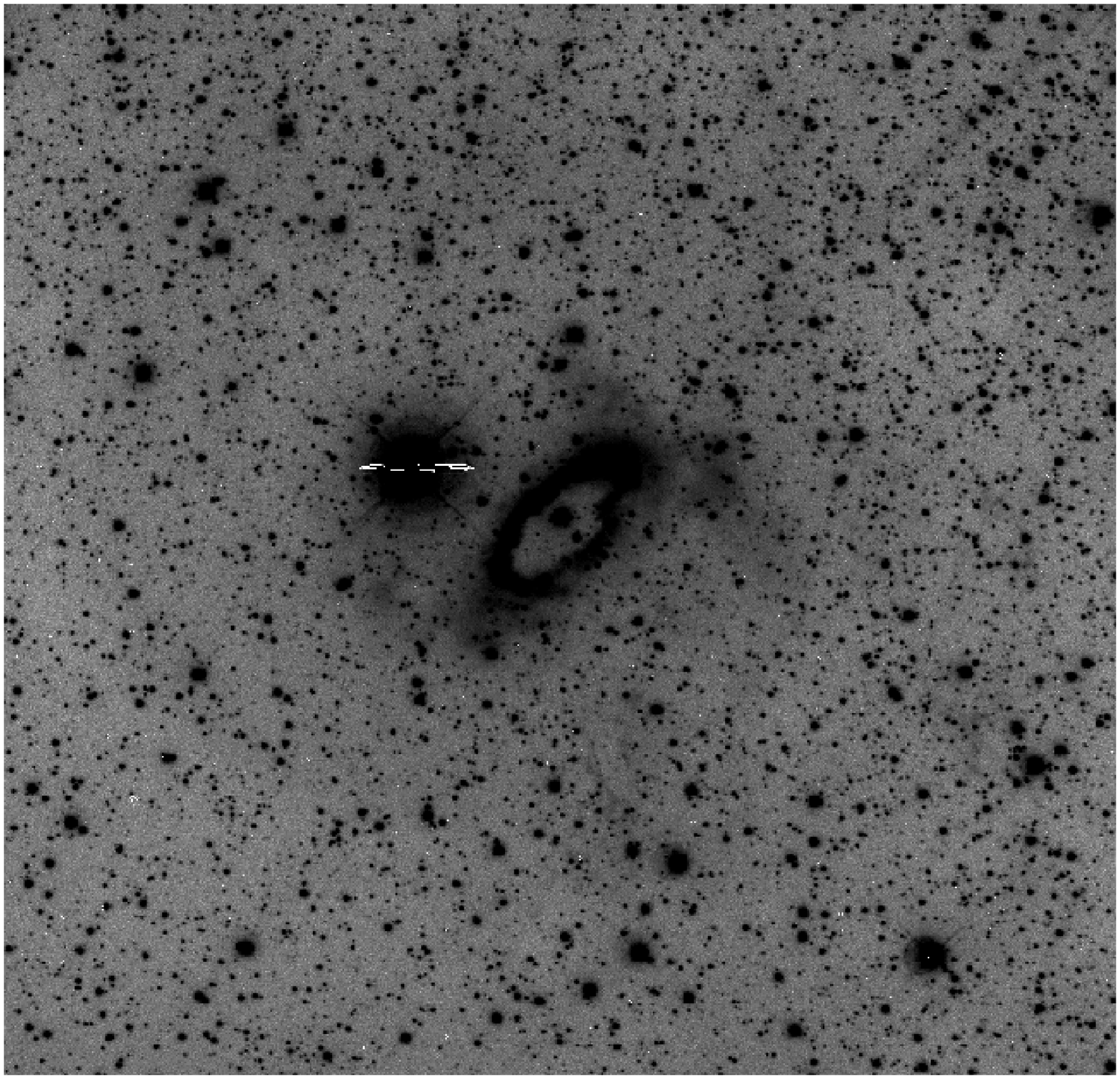}
\caption{CCD image of SuWt~2, obtained with the CTIO 1.5-m telescope using an
H$\alpha$+[\ion{N}{2}] filter and an exposure of $3\times600$~s. A logarithmic
stretch has been used to illustrate the faint lobes extending perpendicularly to
the bright nebular ring. N is up  and E is left, and the image is $8'$ high. The
bright B-type supergiant HD~121228 lies $73''$ ENE of the central star. }
\end{figure}

\begin{figure}[h]
\epsscale{0.6}
\plotone{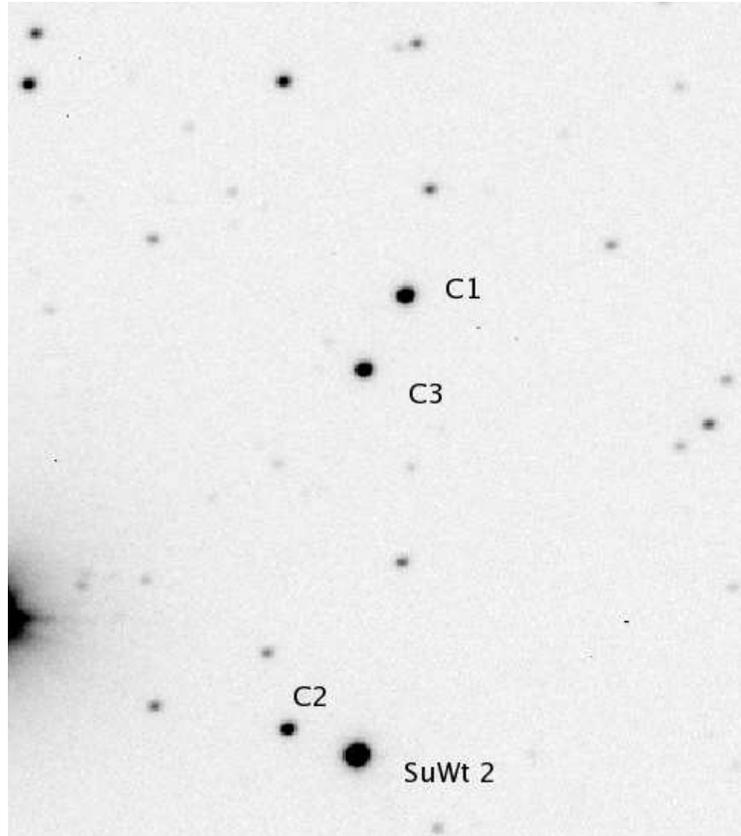}
\caption{Finding chart for central star of SuWt\,2 and its comparison stars,
from a $B$-band frame obtained with the CTIO 0.9-m telescope. N is up and E is
left, and the image is $2\farcm7$ high. }
\label{f:pos}
\end{figure}

\begin{figure}[h]
\epsscale{0.9}
\plotone{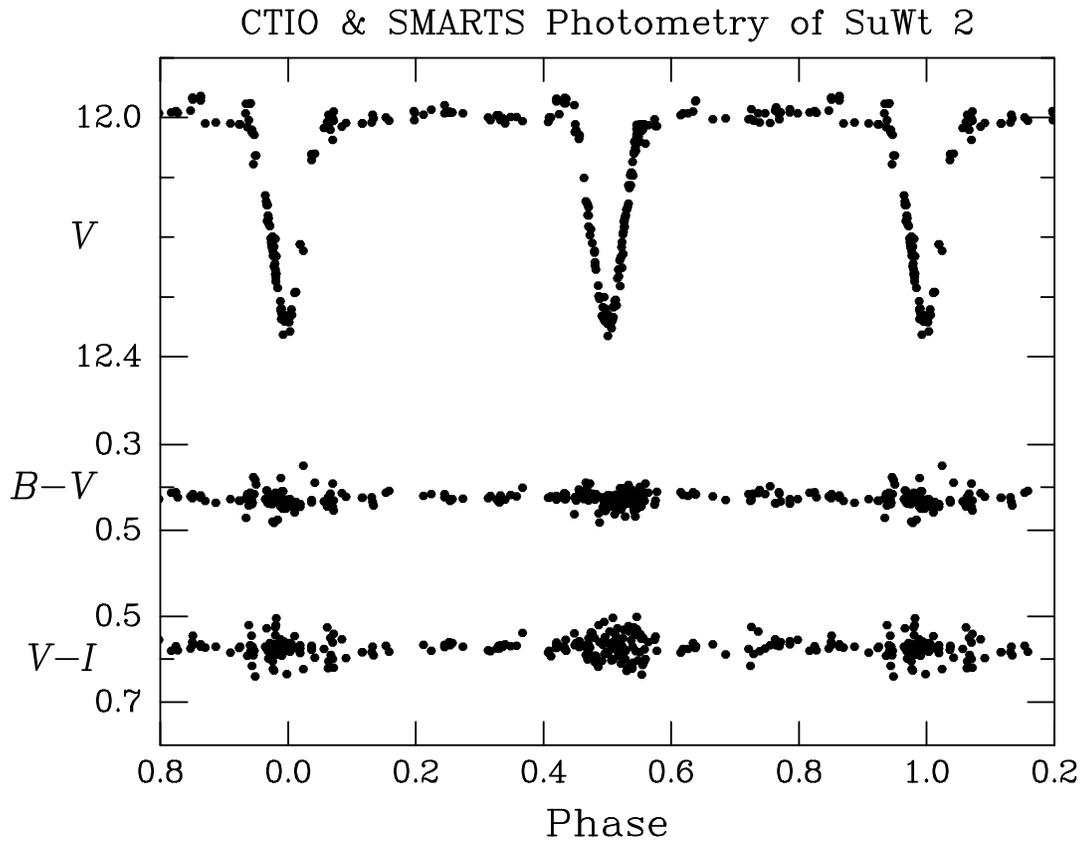}
\caption{$V$-band light curve, and $B-V$ and $V-I$ color curves, of NSV 19992
phased with a period of 4.9098505 days.}
\end{figure}

\begin{figure} 
\label{f:echelle}
\begin{center}
\includegraphics[angle=270,scale=0.45]{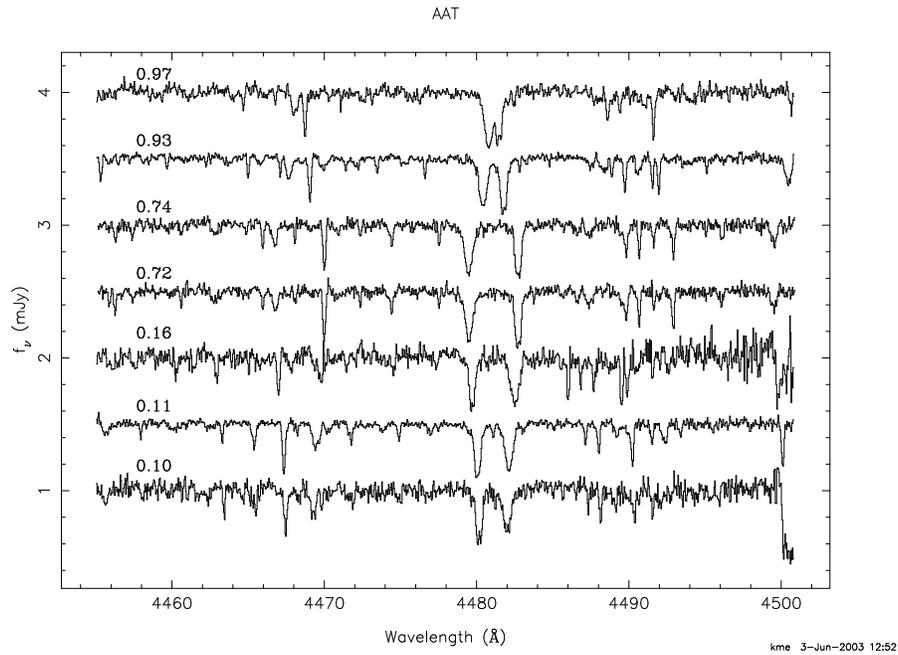}
\end{center}
\caption{Spectra of NSV 19992 taken with the AAT echelle (the phases indicated 
are based on a preliminary ephemeris).  The spectra have been continuum
normalized and shifted successively upwards by 0.5.  Flux is in arbitrary units.
They clearly show the two stars moving in a binary orbit.}
\end{figure}

\begin{figure}[!ht]
\epsscale{0.8}
\plotone{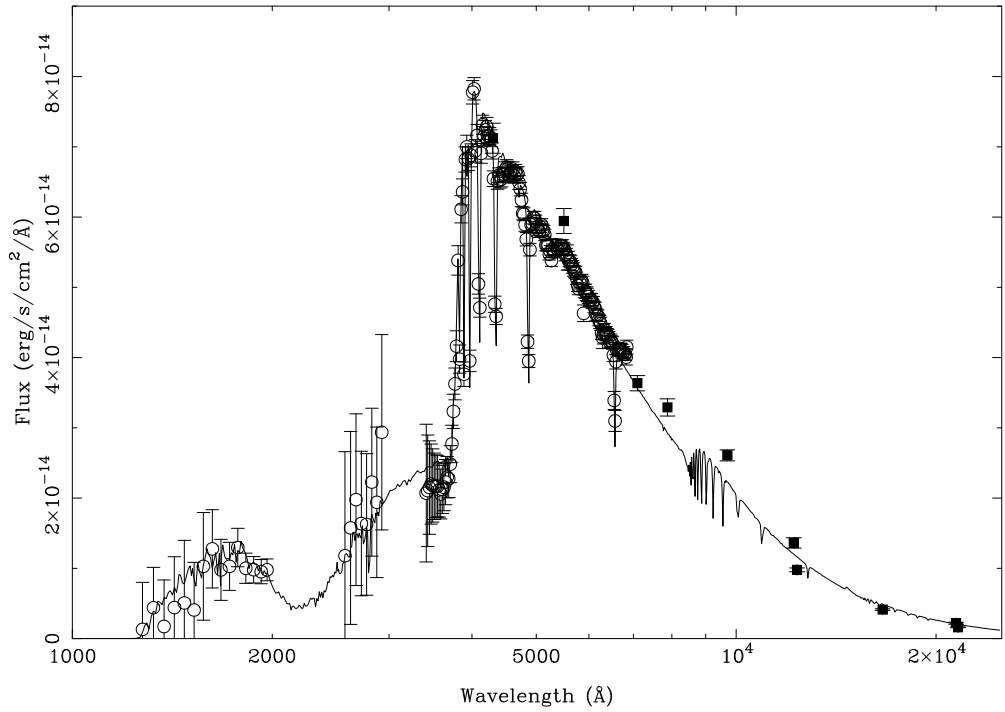}
\caption{The spectral energy distribution, using spectra and photometry from
various sources. Symbols: open--binned {\it IUE\/} and the low-resolution
optical spectrum; filled squares--broad-band photometry (optical, DENIS, 2MASS);
continuous line--best fitting Kurucz model atmosphere flux distribution. 
$E(B-V) = 0.40$.}
\label{flux_plot}
\end{figure}

\begin{figure}[!ht]
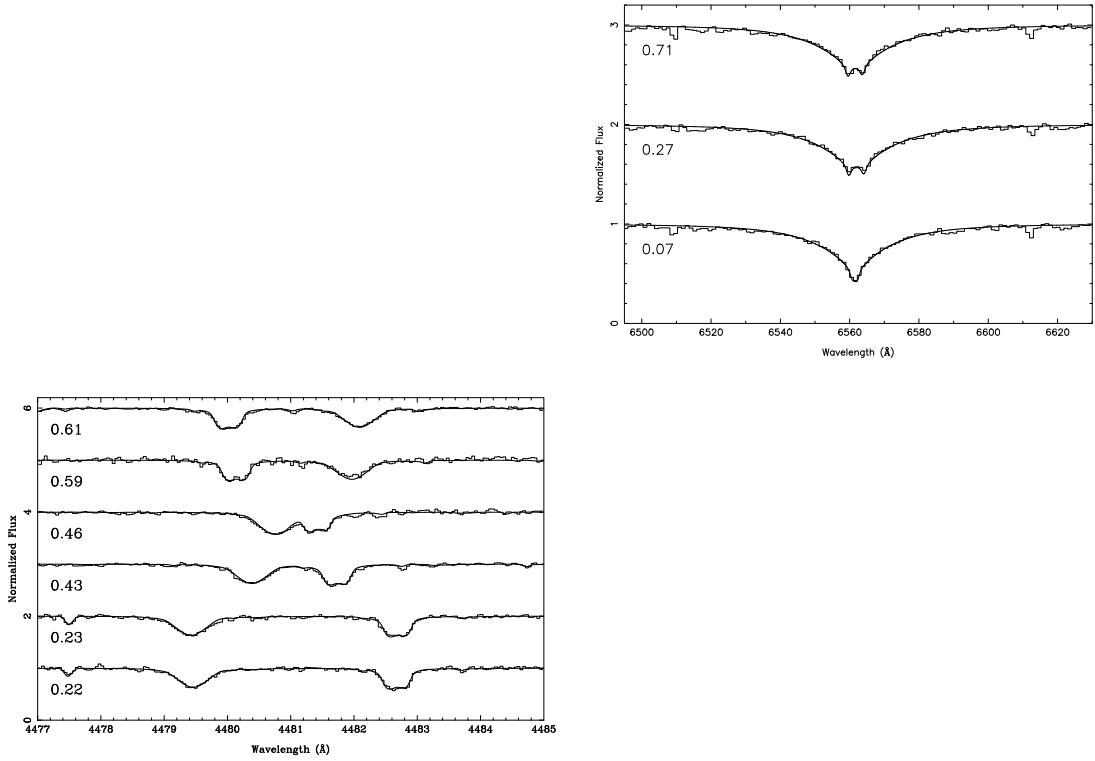

\begin{center}
\includegraphics[angle=-90, scale=0.28]{figure6a.eps}
\includegraphics[scale=0.38]{figure6b.eps}
\end{center}
\caption{The echelle spectra and model fits to the \ion{Mg}{2} 4481~\AA\ region
({\sl left}) and H$\alpha$ ({\sl right}). Histograms are the data and continuous
lines are the synthetic profiles. These profiles have been fitted using the
parameters determined by us; only the radial-velocity values have been allowed
to vary between phases. Phases are marked.}
\label{f:mg_plot}
\end{figure}

\begin{figure}
\label{f:ccf}
\epsscale{1}
\plottwo{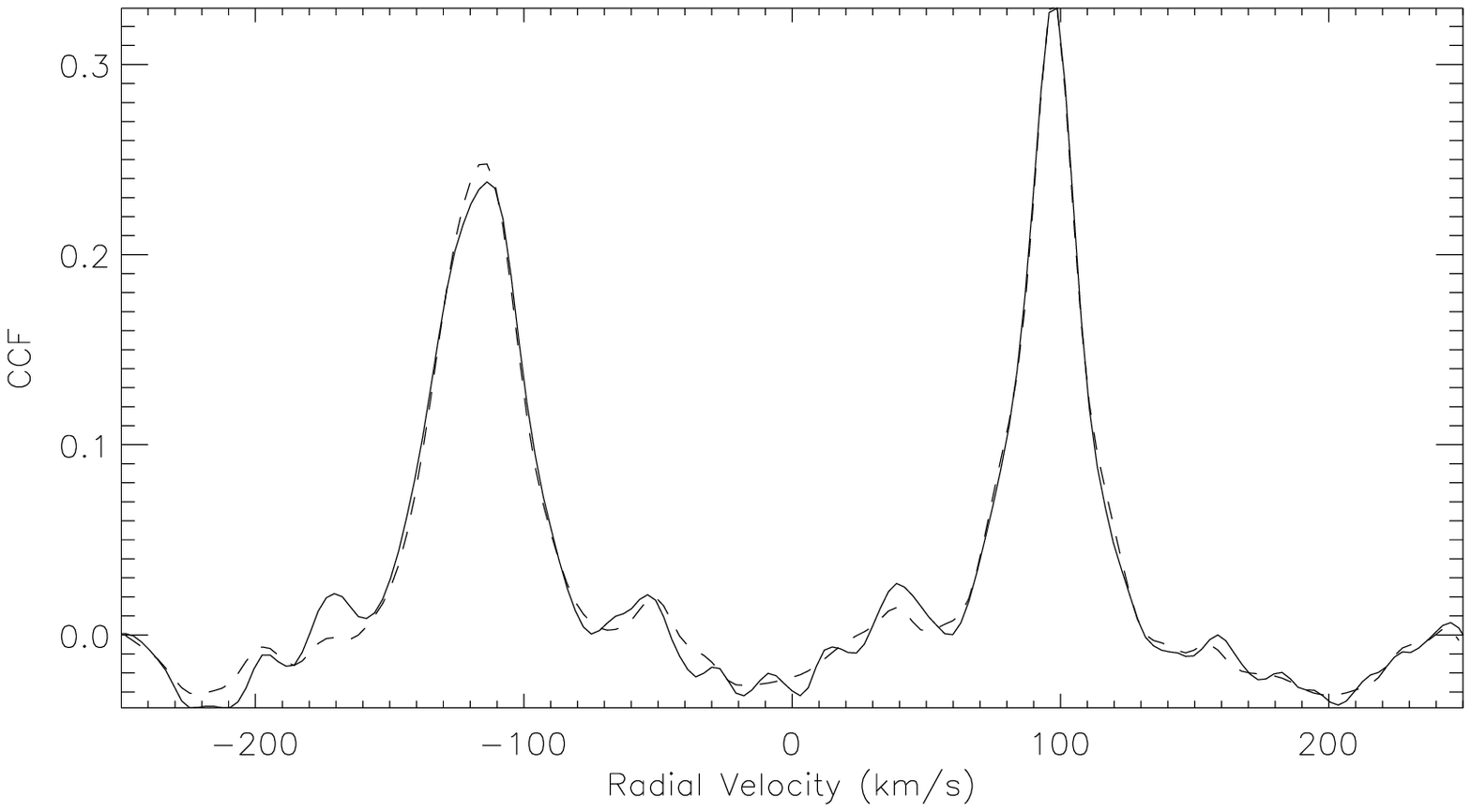}{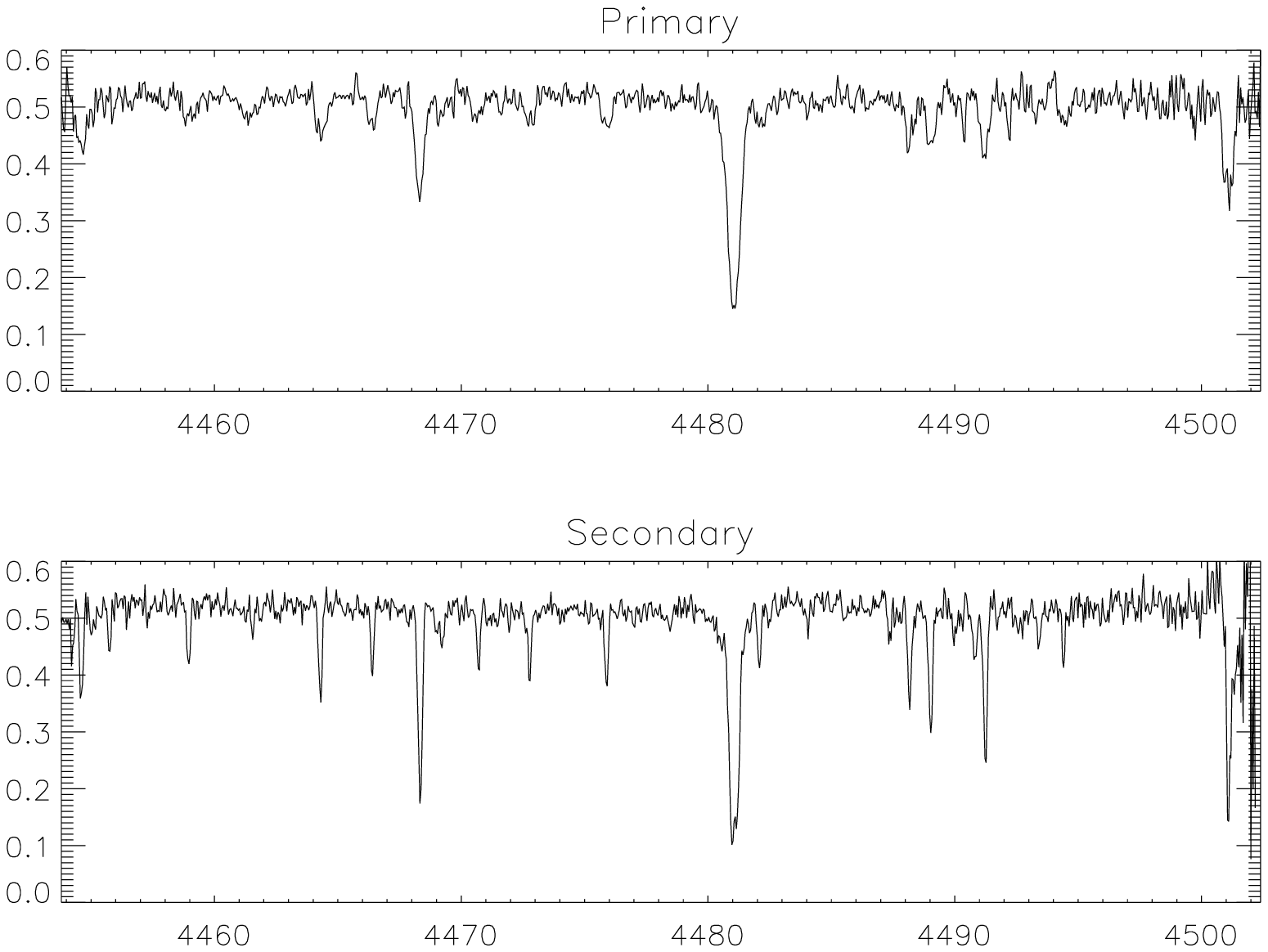}
\caption{{\bf Left} An example of fits to the peaks in the CCFs measured from
one of the model fits to an AAT spectrum. {\bf Right} Disentangled AAT echelle
spectra showing the region around the Mg\,{\sc ii} line(s). Here one can see the
two stars separately, clearly showing the great similarity in the spectra except
for their line widths. }
\end{figure}

\begin{figure}[ht]
\epsscale{0.8}
\plotone{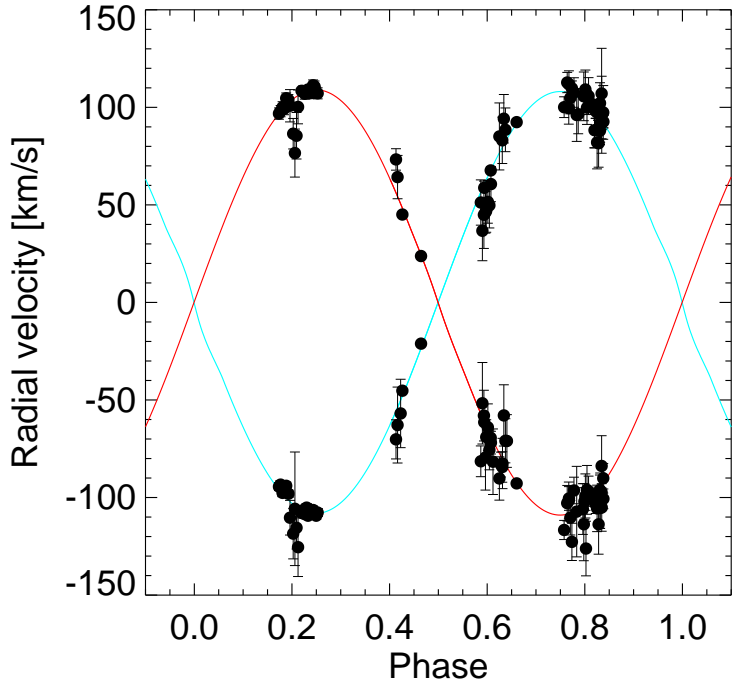}
\caption{\label{fig:rvfitfin}
Orbital fit to the radial-velocity measurements of NSV\,19992. Data from the
different observing runs have been shifted by the systemic velocities listed in
Table~7. Parameters of the orbit solution are summarized in
Tables~8 and 9. The velocity curve of the primary 
component is shown in blue, and that of the secondary in
red.
}
\end{figure}

\begin{figure}[ht]
\begin{center}
\includegraphics[height=5in]{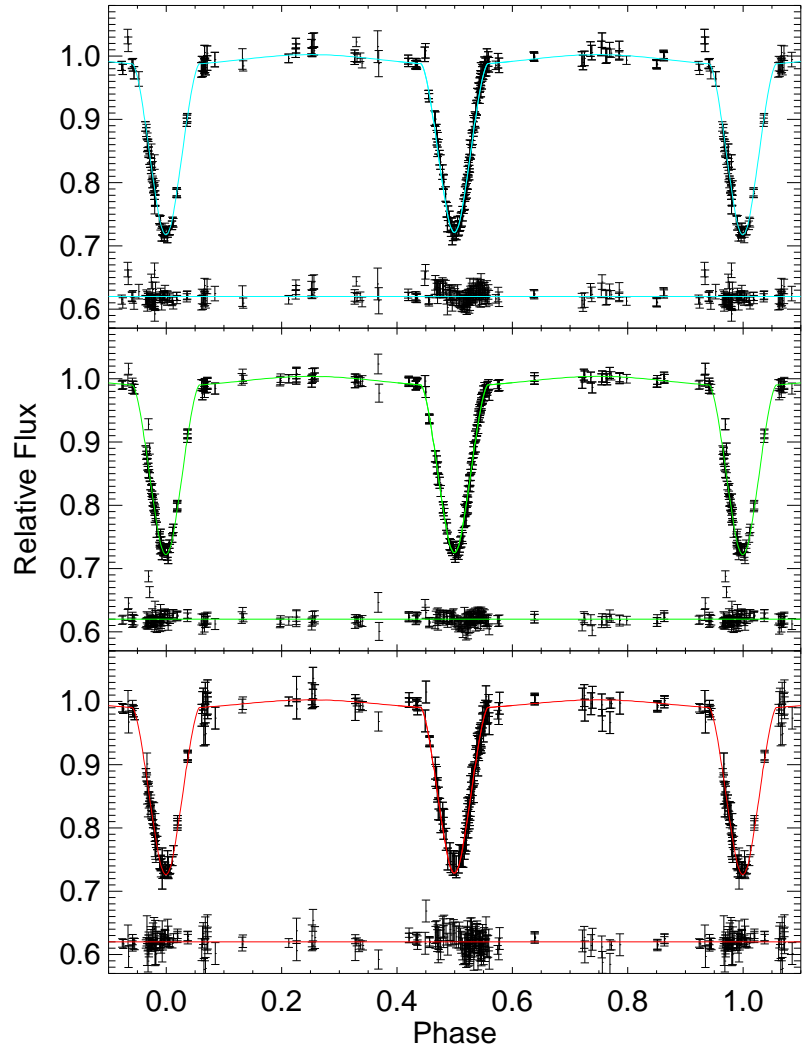}
\end{center}
\caption{\label{fig:lcfit}
Light-curve fits for NSV\,19992. The three panels show the light-curve data and
fit for $B$, $V$, and $I$ bands, respectively. The final physical parameters of
the NSV\,19992 system resulting from this fit are summarized in
Table~9.
}
\end{figure}

\begin{figure*}[ht]
\begin{center}
\includegraphics[width=3.5in]{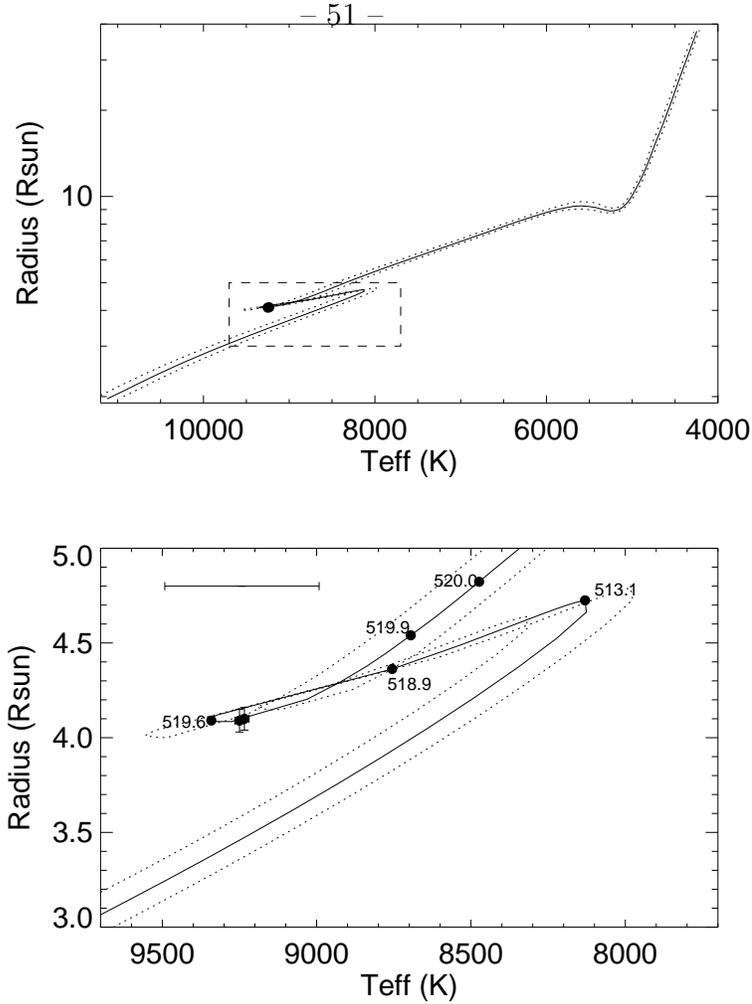}
\end{center}
\caption{\label{fig:radteff}
Evolutionary state of the components of NSV\,19992 is shown in a modified H-R
diagram  ($T_{\rm eff}$ vs.\ $R$). The solid line is an evolutionary track from
the ``Yonsei-Yale" theoretical models for a single star with mass
$2.69\,M_\odot$ and solar metallicity. The dotted curves represent the same
model but for metallicity values of $\rm[Fe/H] \pm 0.14$, representative of the
$1\sigma$ uncertainty in the measured Fe abundance. The measured values for the
components  of NSV\,19992 are shown as filled symbols with error bars. The upper
panel shows an overview of the evolution from the main sequence (lower left
corner) to the base of the red-giant branch.  The lower panel is an expanded
view of the region around the red and blue ``hooks"  (indicated by the dashed
box in the upper panel) representing the transition from core H-burning to shell
H-burning. Stellar ages in Myr from the theoretical model are indicated at
various points along the track. The bar near the top of the lower panel
represents the systematic uncertainty on the stellar $T_{\rm eff}$.}
\end{figure*}

\begin{figure}[ht]
\epsscale{0.9}
\plotone{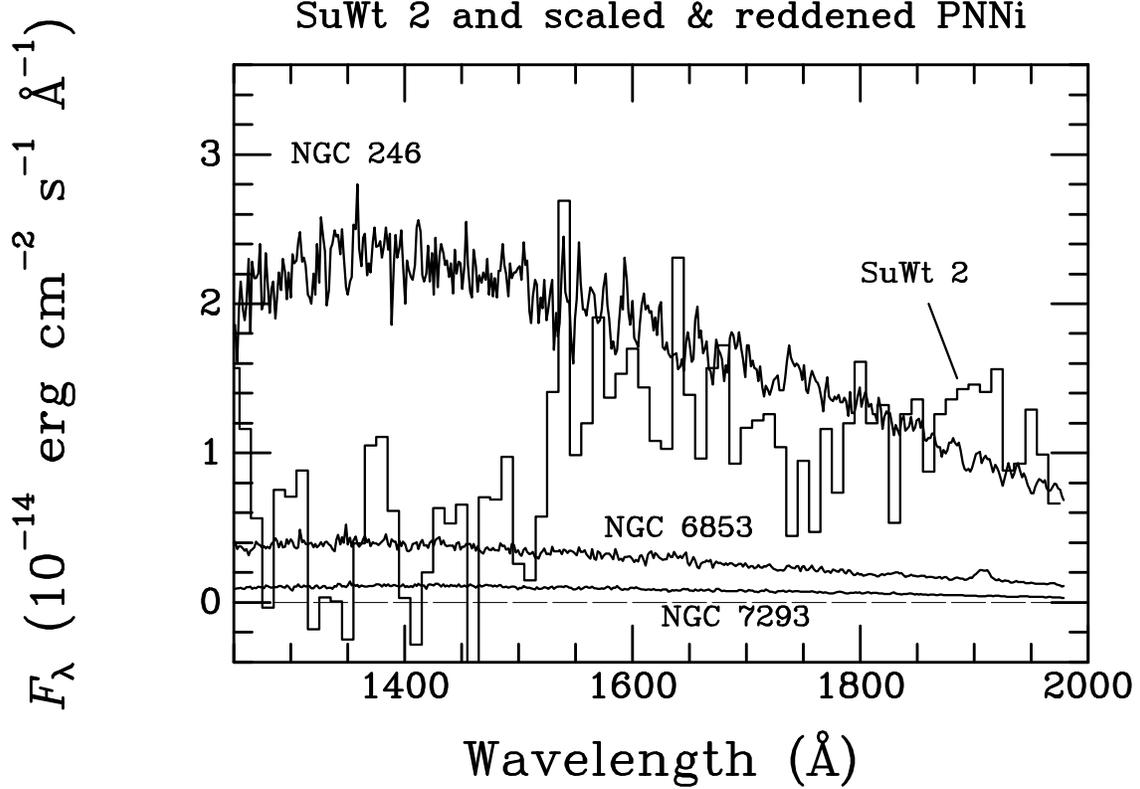}
\caption{\label{fig:iuespectra}
Short-wavelength {\it IUE\/} spectra of NSV\,19992 (histogram; data smoothed to
10~\AA\ bins), NGC~246, NGC~6853, and NGC~7293 (continuous lines). The spectra
of the last three have been scaled from their known distances to 2.3~kpc, and
have also been reddened to a value of $E(B-V)=0.40$. The spectrum of NSV\,19992
is consistent with that of a pair of early A-type stars. Although a PNN as hot
and luminous as that of NGC~246 would have been detected if it were a companion
of the binary, the other two would not have been detected conclusively at the
S/N of the spectrum.
}
\end{figure}


\clearpage


\begin{deluxetable}{lccccc}
\label{t:absphot}
\tabletypesize{\scriptsize}
\tablewidth{0 pt}
\tablecaption{Positions and calibrated photometry for NSV\,19992  outside
eclipse and the comparison stars. Also the DENIS and 2MASS magnitudes of
NSV\,19992}
\tablehead{
\colhead{Star} &
\colhead{$\alpha$ (J2000)} &
\colhead{$\delta$ (J2000)} &
\colhead{$V$} &
\colhead{$B-V$} &
\colhead{$V-I$}
}
\startdata
NSV\,19992 & 13 55 43.23 & $-$59 22 39.8 & $11.999\pm0.006$ & $0.424\pm0.002$ & $0.563\pm0.003$ \\
C1	   & 13 55 42.35 & $-$59 21 15.2 & $12.826\pm0.006$ & $1.377\pm0.004$ & $1.440\pm0.004$ \\
C2	   & 13 55 44.89 & $-$59 22 35.3 & $13.645\pm0.004$ & $1.285\pm0.004$ & $1.408\pm0.004$ \\
C3	   & 13 55 43.31 & $-$59 21 29.0 & $13.742\pm0.004$ & $0.538\pm0.003$ & $0.689\pm0.003$ \\
& & & & & \\
NSV\,19992 & \multicolumn{2}{c}{DENIS mags} & I=$11.405\pm{ 0.04}$ &
J=$10.918\pm{0.06}$ & K=$10.745\pm{0.07}$ \\
NSV\,19992 & \multicolumn{2}{c}{2MASS mags} & J=$11.263\pm{0.024}$&
H=$11.096\pm{0.026}$& K=$11.063\pm{0.025}$ \\
\enddata
\end{deluxetable}

\begin{deluxetable}{lccc}
\label{t:toe}
\tabletypesize{\scriptsize}
\tablewidth{0 pt}
\tablecaption{Eclipse timings of NSV\,19992\label{tab:toe}}
\tablehead{
\colhead{HJD (2450000+)} &
\colhead{$E$} &
\colhead{$(O-C)$ [phase]} & 
\colhead{Filter}
}
\startdata
 $ 599.8417 \pm 0.0013$ & $-14$ & $-0.0024$ & $B$ \\
 $ 599.8475 \pm 0.0011$ & $-14$ & $-0.0012$ & $V$ \\
 $ 599.8479 \pm 0.0012$ & $-14$ & $-0.0012$ & $I$ \\
 $ 668.5857 \pm 0.0018$ & 0	& $-0.0012$ & $V$ \\
 $ 668.5868 \pm 0.0027$ & 0	& $-0.0010$ & $I$ \\
 $ 668.5907 \pm 0.0017$ & 0	& $-0.0002$ & $B$ \\
 $ 923.9079 \pm 0.0029$ & 52	& $+0.0008$ & $B$ \\
 $ 923.9085 \pm 0.0018$ & 52	& $+0.0010$ & $V$ \\
 $1051.5509 \pm 0.0040$ & 78	& $-0.0018$ & $B$ \\
 $1051.5611 \pm 0.0068$ & 78	& $+0.0003$ & $I$ \\
 $1051.5648 \pm 0.0029$ & 78	& $+0.0010$ & $V$ \\
 $1343.6958 \pm 0.0028$ & 137.5 & $-0.0000$ & $B$ \\
 $1343.6959 \pm 0.0023$ & 137.5 & $-0.0000$ & $V$ \\
 $1343.6974 \pm 0.0059$ & 137.5 & $+0.0003$ & $I$ \\
 $1991.8001 \pm 0.0012$ & 269.5 & $+0.0008$ & $V$ \\
 $1991.8016 \pm 0.0026$ & 269.5 & $+0.0011$ & $I$ \\
 $1991.8029 \pm 0.0015$ & 269.5 & $+0.0014$ & $B$ \\
 $4208.5838 \pm 0.0029$ & 721	& $-0.0020$ & $I$ \\
 $4208.5861 \pm 0.0023$ & 721	& $-0.0016$ & $V$ \\
 $4208.5925 \pm 0.0021$ & 721	& $-0.0003$ & $B$ \\
\enddata
\end{deluxetable}

\begin{deluxetable}{lccc}
\label{t:phot}
\tabletypesize{\scriptsize}
\tablewidth{0 pt}
\tablecaption{Details of the spectroscopic observing runs}
\tablehead{
\colhead{Date} &
\colhead{Telescope}&
\colhead{Wavelength Calbration} &
\colhead{Exposure} \\
\colhead{month/day[range]/year} &
\colhead{}&
\colhead{Range/Dispersion/RMS} &
\colhead{} \\
}
\startdata
06/13/90 & $IUE$ & 1150--1970\,\AA; low  & 4200s \\
06/25/91 & $IUE$ & 1150--1970\,\AA; low   & 13500s \\
06/25/91& $IUE$ &1850--3350\,\AA;low & 1500s  \\
05/23--29/95 & SAAO 1.9m (RPCS) & 4080--5160\,\AA;0.8\,\AA;0.05--0.1\,\AA\,pix$^{-1}$  &1200s  \\
05/21--23/00 & AAO AAT  (UCLES) & 4455--4500\,\AA;0.12\,\AA;--\tablenotemark{a} &1800s\\
04/3--5/01 & AAO AAT  (RGO) & 4138--4577\,\AA;0.55\,\AA;0.02--0.03\,\AA\,pix$^{-1}$ & 1200s  \\
04/19--22/95 & ESO NTT (EMMI) & 4080--5020\,\AA;2.7\,\AA;0.02--0.03\,\AA\,pix$^{-1}$ & 900s\\
03/[24,27,29]/01; 02/[13,15,16,21]/03 & CTIO 1.5m  (RC Spectrograph) & 4019--4931\,\AA;1.6\,\AA;0.04\,\AA\,pix$^{-1}$ & 300--400s  \\
03/13/03; 06/11/06 & CTIO 1.5m  (RC Spectrograph) & 4019--4931\,\AA;1.6\,\AA;0.04\,\AA\,pix$^{-1}$ & 300--400s  \\
03/[24,27,29]/01; 02/[13,15,16,21]/03& CTIO 1.5m  (RC Spectrograph) & 6012--7332\,\AA;3.1\,\AA;0.04\,\AA\,pix$^{-1}$ &  60s   \\
03/13/03; 06/11/06  & CTIO 1.5m  (RC Spectrograph) & 6012--7332\,\AA;3.1\,\AA;0.04\,\AA\,pix$^{-1}$ &  60s   \\
01/21/07 & CTIO 1.5m (RC Spectrograph) & 3400--6850\,\AA;11.6\,\AA;0.12\,\AA\,pix$^{-1}$& 300s\\
\enddata
\tablenotetext{a}{RMS was not recorded}
\end{deluxetable}

\begin{deluxetable}{lll}
\label{tab:stellparams}
\tablewidth{0 pt}
\tablecaption{Stellar parameters and abundances from spectral synthesis}
\tablehead{
\colhead{Quantity} &
\colhead{Primary Star}  &
\colhead{Secondary Star}
}
\startdata
\teff\ (K)          & 9250 $\pm$ 250    & 9150 $\pm$ 250   \\
\logg               & 4.0 $\pm$ 0.3     & 3.5 $\pm$ 0.3    \\
{[Fe/H]}            & +0.01 $\pm$ 0.14  & 0.00 $\pm$ 0.14  \\
{[Ti/H]}            & +0.59 $\pm$ 0.15  & +0.44 $\pm$ 0.16 \\
{[Mg/H]}            & +0.34 $\pm$ 0.20  & +0.41 $\pm$ 0.20 \\ 
\vsini\ (\kms)      & 17 $\pm$ 3        & $<$5             \\
\mictrb\ (\kms)     & 3.2 $\pm$ 0.2     & 2.6 $\pm$ 0.2    \\
Luminosity fraction & 0.50$\pm$0.02     & 0.50$\pm$0.02    \\
\enddata
\end{deluxetable}

\begin{deluxetable}{lccc|lccc}
\label{t:rv}
\tabletypesize{\scriptsize}
\tablewidth{0 pt}
\tablecaption{Radial velocities (km\,s$^{-1}$)}
\tablehead{
\colhead{HJD}&
\colhead{Secondary} &
\colhead{Primary} &
\colhead{Phase} &
\colhead{HJD}&
\colhead{Secondary} &
\colhead{Primary}&
\colhead{Phase} \\
\colhead{(24+)}&
\colhead{} &
\colhead{} &
\colhead{} &
\colhead{(24+)}&
\colhead{} &
\colhead{}&
\colhead{} 
}
\startdata
\multicolumn{4}{c}{AAT echelle UCLES}                             &  52005.18208 & $102.9\pm2.7$&  $-113.7\pm2.9$      & .226 \\     
 51686.00921 &   $100.66\pm0.12$&  $-115.32\pm0.21     $ & .219 & 52005.20053 & $103.5\pm2.9$ &  $-110.4\pm3.0$  & .230 \\      
 51686.07438 &   $101.32\pm0.14$&  $-116.75\pm0.22     $ & .232 & 52005.21536 & $103.9\pm2.9$  &  $-111.1\pm3.1$ & .233 \\       
 51687.02352 &   $37.24 \pm0.14$&  $-52.69\pm0.24       $& .426 & 52005.23149 & $103.2\pm2.7$  &  $-112.3\pm2.8$ & .236 \\       
 51687.21123 &   $15.96 \pm0.26$&  $-28.56\pm0.35      $ & .464 &52005.24632 &  $104.6\pm2.8$  &  $-112.3\pm3.0$ & .239 \\      
 51687.84689 &  $-69.21\pm0.25$&   $51.33 \pm0.43      $ & .593 & 52005.26550 & $107.2\pm2.8$  &  $-111.6\pm2.9$ & .243 \\      
 51687.91204 &  $-77.22\pm0.16$&   $60.24\pm0.25     $   & .607 &52005.28034 &  $106.6\pm2.8$  &  $-112.9\pm2.9$ & .246 \\      
 51688.17285 &  $-100.57 \pm0.24$&   $84.98\pm0.38$      & .660 &52005.29655 &  $103.4\pm2.8$  &  $-114.5\pm2.9$ & .249 \\      
\multicolumn{4}{c}{AAT mid-resolution RGO}                      &   52005.31138 & $103.1\pm2.8$ &  $-112.9\pm3.0$    & .252 \\      
 52003.09624 &  $ -104.9\pm5.8  $  &                     & .801 &  \multicolumn{4}{c}{NTT low resolution EMMI }\\                     
 52003.16759 &  $ -104.9\pm3.0 $   &   $96.3\pm5.0$      & .816 &  49827.817370& $-159.7\pm4.9$& $58.6\pm5.3$    & .757 \\         
 52003.18666 &                    &  $ 83.1\pm8.0$       & .819 &  49827.865438&$-143.5\pm6.1$ &$70.3\pm6.9$     & .767 \\                   
 52003.19897 &   $-107.6\pm4.6   $ & $  92.8\pm4.6 $     & .822 &  49827.916331&$-139.4\pm6.8$ &$66.1\pm7.6$     & .777 \\                
 52003.22556 &   $-103.2\pm4.4   $ &  $ 91.5\pm4.4 $     & .827 &  \multicolumn{4}{c}{SAAO mid-resolution RPCS}\\                 
 52003.24120 &   $-100.1\pm7.6   $ &   $82.8\pm8.1$      & .831 &     49861.40278 &$-90.4\pm8.9 $ & $30.0\pm11.1$& .598\\      
 52003.25740 &   $-101.4\pm2.2   $ &   $86.6\pm4.2$      & .834 &     49861.41874 &$-85.4\pm12.0 $ &$34.5\pm10.9$& .601\\    
 52003.27586 &   $-104.7\pm5.0   $ &   $87.4\pm4.1$      & .838 &   49861.43428 &$-97.0\pm10.2$  &$32.7\pm11.6 $ & .604\\
 52004.92234 &    $92.7\pm2.9  $ & $ -99.7\pm3.0$        & .173 &   49861.44994 &$-93.2\pm5.8 $  &  $43.6\pm9.0$ & .607\\  
 52004.93856 &    $94.0\pm2.9  $ &  $-98.6\pm3.0$        & .176 &  49861.56414 &$-105.7\pm7.8  $ & $66.0\pm1.9 $ & .630\\  
   52004.95863 &    $96.0\pm2.8  $ &  $-102.9\pm2.9$     & .180 &   49861.57973 &$-79.3\pm15.7 $ & $77.1\pm12.4 $& .634\\   
       52004.97347 &    $95.1\pm3.0 $  &  $-99.9\pm3.2  $& .183 &  49861.59529 &$-92.3\pm11.3 $ & $71.4\pm11.2$  & .637\\  
 52004.99521 &    $100.5\pm2.7$   &  $-99.2\pm2.8$       & .188 &  49861.61082 &$-92.4\pm13.5 $ &                & .640\\  
 52005.01723 &    $99.8\pm2.6$   &  $-103.3\pm2.8 $      & .192 &  49862.21732 &$-124.2\pm9.1  $ &  $95.6\pm5.2$ & .764\\ 
 52005.16725 &    $104.3\pm2.7$   &  $-111.6\pm2.8$      & .223 &  49862.23307 &$-122.8\pm9.42$ &  $ 82.5\pm8.2$ & .767\\              
 \enddata
\end{deluxetable}

\addtocounter{table}{-1}  
\begin{deluxetable}{lccc|lccc}
\label{t:rv}
\tabletypesize{\scriptsize}
\tablewidth{0 pt}
\tablecaption{Radial velocities (km\,s$^{-1}$); con't}
\tablehead{
\colhead{HJD}&
\colhead{Secondary} &
\colhead{Primary}&
\colhead{Phase} &
\colhead{HJD}&
\colhead{Secondary} &
\colhead{Primary}&
\colhead{Phase}\\
\colhead{(24+)}&
\colhead{} &
\colhead{} &
\colhead{} &
\colhead{(24+)}&
\colhead{} &
\colhead{}&
\colhead{}
}
\startdata
\multicolumn{7}{c}{SAAO mid-resolution RPCS}\\
   49862.24861 &$-131.7\pm8.87 $ &    $ 87.7\pm8.1   $ & .770 & 49867.22227 &$-128.6\pm23.2$ &  $   79.0\pm13.5 $  & .783\\                 
   49862.26419 &$-144.2\pm9.5 $ &    $ 92.7\pm3.8    $ & .773 & 49867.23781 &           & $  79.4\pm10.0  $        & .786\\                    
   49862.38420 &$-135.2\pm18.7 $&   $  88.6\pm12.5 $   & .797 & 49867.28447 &$-127.3\pm13.0$            &          & .796\\                 
   49862.39977 &$-120.5\pm11.3 $&   $  92.1\pm10.0   $ & .801 & 49867.30002 &$-123.4\pm9.2  $           &          & .799\\                         
 49862.41619 &$-117.4\pm12.5 $&   $  82.8\pm8.6   $    & .804 &  49867.31561 &$-147.5\pm14.0 $           &         & .802\\               
   49862.43180 &$-118.1\pm7.7 $ &    $ 88.7\pm9.5    $ & .807 & 49867.42737 &$-126.9\pm11.9 $ &   $  65.0\pm13.6$  & .825\\               
   49862.54706 &$-126.7\pm10.5$ &   $  85.0\pm10.5  $  & .831 &   49867.44294 &$-135.1\pm15.3$ &    $ 64.5\pm12.3 $& .828\\           
   49862.56327 &$-105.2\pm15.5$ &   $  89.0\pm23.2 $   & .834 & 49867.45848 &$-122.9\pm14.2 $ &  $   74.2\pm9.2  $ & .831\\           
   49862.57945 &$-111.5\pm7.6 $ &    $ 80.3\pm13.8  $  & .837 &  49867.47402 &$-126.4\pm12.2$ &  $   79.1\pm19.7  $& .834\\            
 49866.25795 &$-102.8\pm7.9 $ &    $ 34.1\pm11.6  $    & .586 &  49864.33839 & $79.5\pm8.3 $ &    $-127.4\pm8.9   $& .196\\           
   49866.27357 &$-73.0\pm20.9$ &    $ 19.7\pm15.3 $    & .590 &   49864.37070 & $65.1\pm7.8 $ &    $-135.6\pm12.9$ & .202\\           
   49866.28914 &$-79.4\pm13.0$ &    $ 27.8\pm17.3$     & .593 & 49864.38793 & $55.2\pm12.3 $&  $  -122.8\pm29.1  $ & .206\\             
   49866.30470 &$-100.8\pm16.7$ &     $29.1\pm10.7 $   & .596 &  49864.40409 & $64.1\pm11.9 $ &  $  -132.6\pm10.9$ & .209\\          
   49866.38020 &$-103.0\pm16.8$           &            & .611 &       49864.41963 & $78.9\pm8.7 $ & $-142.5\pm15.0$& .212\\          
   49866.44475 &$-111.6\pm11.1$ &   $  68.0\pm17.2$    & .625 &       49865.40475 & $51.9\pm5.5 $ & $ -87.2\pm10.0$& .413\\            
   49866.46029 &$-104.5\pm6.4 $ &                      & .628 &  49865.42029 & $42.9\pm11.1$ & $   -79.9\pm19.5  $ & .416\\             
   49866.47583 &$-103.9\pm12.9$ &                      & .631 & 49865.45146 &            &   $-74.0\pm17.5     $   & .422\\
    \enddata                                                                                               
\end{deluxetable}

\begin{deluxetable}{lc}
\label{t:rvfitaat}
\tablewidth{0 pt}
\tablecaption{Orbital parameters of NSV\,19992 from AAT echelle data only}
\tablehead{
\colhead{Parameter} &
\colhead{Value}
}
\startdata 
$q \equiv M_2 / M_1\tablenotemark{a}$ & $0.9921\pm0.0097$ \\
$a \sin i \, [{\rm R}_\odot]$ & $21.096\pm0.147$ \\
$e$ & $0.992\pm0.004$  \\
$M_{\rm tot} \sin^3 i \, [{\rm M}_\odot]$ & $5.232\pm0.070$ \\
$K_1$ & $109.37\pm{0.50}$ km s$^{-1}$ \\ 
$K_2$ & $110.23\pm{0.50}$ km s$^{-1}$ \\
$M_1 \sin^3i$ &$2.616\pm{0.037} {\rm M}_\odot$ \\ 
$M_2 \sin^3i$ &$2.596\pm{0.037} {\rm M}_\odot$ \\ 
P & $4.9098505\pm{0.0000020}$ d \\
T$_0$ &$2450668.5915\pm{0.0010}$ d\\
\enddata
\tablenotetext{a}{1=primary; 2=secondary}
\end{deluxetable}

\begin{deluxetable}{lcc}
\label{tab:vgamma}
\tablewidth{0 pt}
\tablecaption{Systemic Velocity of NSV\,19992 from each Epoch of Radial-Velocity
Data}
\tablehead{
\colhead{Observing Run} &
\colhead{Mean Epoch} &
\colhead{$v_\gamma$ [$\kms$] }
}
\startdata 
NTT low res.  & 1995.30 & $-43.08\pm2.72$ \\
SAAO mid res. & 1995.40 & $-21.36\pm1.13$ \\
AAT echelle   & 2000.39 & $-7.82\pm0.41$ \\
AAT mid res.  & 2001.25 & $-4.08\pm0.36$ \\
\enddata
\end{deluxetable}

\begin{deluxetable}{lcc}
\label{tab:rvfitfin}
\tablewidth{0 pt}
\tablecaption{Orbital parameters of NSV\,19992 from fit to all radial velocity data}
\tablehead{
\colhead{Parameter} &
\colhead{Value} 
}
\startdata 
$q \equiv M_2 / M_1$& $0.992\pm0.004$ \\
$a \sin i \, [{\rm R}_\odot]$ & $21.066\pm0.050$ \\
$e$ & $0.023\pm0.011$\tablenotemark{a} \\
$M_{\rm tot} \sin^3 i \, [{\rm M}_\odot]$ & $5.210\pm0.037$ \\
    \enddata
\tablenotetext{a}{Eccentricity fixed at zero for calculations--see \S6.2}
\end{deluxetable}

\begin{deluxetable}{lcc}
\label{tab:finparams}
\tablewidth{0 pt}
\tablecaption{Physical parameters of the components of
NSV\,19992}
\tablehead{
\colhead{Parameter} &
\colhead{Value} 
}
\startdata 
$i \, [^\circ]$          & $81.5\pm0.1$ \\
$M_1 \, [M_\odot]$    & $2.704\pm 0.038$ \\
$M_2 \, [M_\odot]$       & $2.683\pm0.038$ \\
$T_1 / T_2$              & $1.0016\pm0.0018$ \\
$R_1 + R_2 \, [R_\odot]$ & $8.195\pm0.046$ \\
$R_1 / R_2$\tablenotemark{a}            & $0.996\pm0.029$ \\
    \enddata
\tablenotetext{a}{Ratio of the radii determined from the temperature and the
spectroscopically determined luminosity ratio $L_1 / L_2 = 1.000\pm0.057$ }
\end{deluxetable}

\end{document}